\documentclass[useAMS,usegraphicx]{mn2e}
\usepackage{amssymb}
\usepackage{lscape}
\usepackage{pifont}
\usepackage{amsfonts}
\usepackage{amsmath}
\usepackage{rotating}
\usepackage{stfloats}
\usepackage{longtable,lscape}

\def\kms{km${\rm s}^{-1}$}
\def\cms2{cm${\rm s}^{-2}$}
\def\arcdeg{\hbox{$^\circ$}}
\def\arcmin{\hbox{$^\prime$}}
\def\arcsec{\hbox{$^{\prime\prime}$}}
\def\ergcms{erg\,cm$^{-2}$\,s$^{-1}$}

\def\ha{H$\alpha$}
\def\hb{H$\beta$}
\def\hg{H$\gamma$}
\def\NII{[N\,\textsc{ii}]}
\def\SII{[S\,\textsc{ii}]}
\def\SIII{[S\,\textsc{iii}]}

\def\OIII{[O\,\textsc{iii}]}

\def\HII{H\,\textsc{ii}}

\def\HeII{He\,\textsc{ii}}
\def\p0{\phantom{0}}

\def\lessim{\raise-.5ex\hbox{$\buildrel<\over{\scriptstyle\mathtt{\sim}}$}}
\def\grtsim{\raise-.5ex\hbox{$\buildrel>\over{\scriptstyle\mathtt{\sim}}$}}

\title[Integrated H$\alpha$ fluxes for Galactic planetary nebulae]{A catalogue of integrated H$\alpha$ fluxes for 1,258 Galactic planetary nebulae}
\author[D.J. Frew, I.S. Boji\v{c}i\'c and Q.A. Parker]
{David J. Frew$^{1,2}$\thanks{E-mail: david.frew@mq.edu.au}, Ivan S. Boji\v{c}i\'c$^{1,2,3}$ and Q.A. Parker$^{1,2,3}$
\\
$^{1}$Department of Physics and Astronomy, Macquarie University, NSW 2109, Australia\\
$^{2}$Research Centre in Astronomy, Astrophysics and Astrophotonics, Macquarie University, NSW 2109, Australia\\
$^{3}$Australian Astronomical Observatory, PO Box 915, North Ryde, NSW 1670, Australia}

\begin{document}

\date{Accepted ; Received ; in original form }
\pagerange{\pageref{firstpage}--\pageref{lastpage}} \pubyear{2012}

\maketitle
\label{firstpage}

\begin{abstract}

We present a catalogue of new integrated \ha\ fluxes for 1258 Galactic planetary nebulae (PNe), with the majority, totalling 1234, measured from the Southern H$\alpha$ Sky Survey Atlas (SHASSA) and/or the Virginia Tech Spectral-line Survey (VTSS).  Aperture photometry on the continuum-subtracted digital images was performed to extract \ha+\NII\ fluxes in the case of SHASSA, and \ha\ fluxes from VTSS. The  \NII\ contribution was then deconvolved from the SHASSA flux using spectrophotometric data taken from the literature or derived by us.  Comparison with previous work shows that the flux scale presented here has no significant zero-point error.   Our catalogue is the largest compilation of homogeneously derived PN fluxes in any waveband yet measured, and will be an important legacy and fresh benchmark for the community.   Amongst its many applications, it can be used to determine statistical distances for these PNe, determine new absolute magnitudes for delineating the faint end of the PN luminosity function, provide baseline data for photoionization and hydrodynamical modelling, and allow better estimates of Zanstra temperatures for PN central stars with accurate optical photometry.  We also provide total \ha\ fluxes for another 76 objects which were formerly classified as PNe, as well as independent reddening determinations for $\sim$270~PNe, derived from a comparison of our \ha\ data with the best literature \hb\ fluxes.   In an appendix, we list corrected \ha\ fluxes for 49 PNe taken from the literature, including 24 PNe not detected on SHASSA or VTSS, re-calibrated to a common zero-point.
\end{abstract}


\section{Introduction}

Planetary nebulae (PNe) are a key end-point in the evolution of mid-mass stars which range from $\sim$1 to 8 times the mass of the Sun.  While the number of known Galactic PNe has essentially doubled over the last decade (Parker et al. 2012a), many hundreds of PNe have very little observational data such as an integrated Balmer-line flux.  The integrated flux is analogous to the apparent magnitude of a star, and is one of the most fundamental observable parameters that needs to be determined for any PN.  Calculations involving the distance, the ionized mass and electron density, the temperature and luminosity of the central star, and the PN luminosity function (PNLF; Ciardullo 2010) are all critically dependent on accurate integrated line fluxes.  Indeed, the majority of distance estimates for Galactic PNe are determined from a statistical method, which depend on having an accurate integrated flux either in the radio continuum (e.g. Milne \& Aller 1975; Daub 1982; Cahn, Kaler \& Stanghellini 1992; van de Steene \& Zijlstra 1994; Zhang 1995; Bensby \& Lundstr\"om 2001; Phillips 2004; Stanghellini, Shaw \& Villaver 2008), or in an optical Balmer line (O'Dell 1962; Frew \& Parker 2006, 2007; Frew 2008).  


Traditionally, the \hb\ flux has been determined, based on the historical precedence of blue-sensitive photoelectric photometers that were widely used from the 1950s to the 1980s.  Integrated \hb\ and \OIII\ fluxes for PNe were measured either with an objective prism (Liller \& Aller 1954), with scanning spectrographs (Capriotti \& Daub 1960; Collins, Daub \& O'Dell 1961; Liller \& Aller 1963; Aller \& Faulkner 1964; O'Dell \& Terzian 1970; Peimbert \& Torres-Peimbert 1971; Barker \& Cudworth 1984) or using conventional aperture photometry with interference filters (e.g.  Liller 1955;  Osterbrock \& Stockhausen 1961; Collins, Daub \& O'Dell 1961; O'Dell 1962, 1963; Webster 1969; Perek 1971; Kaler 1976, 1978; Carrasco, Serrano \& Costero 1983, 1984; Shaw \& Kaler 1985).  Consequently, most of the brighter Galactic PNe now have integrated  \hb\ fluxes available, as compiled by Acker et al. (1991, hereafter ASTR91) and Cahn, Kaler \& Stanghellini (1992, hereafter CKS92), based on the efforts of many observers over several decades (see references therein).  A few studies (e.g. Webster 1969, 1983; Kohoutek \& Martin 1981a, hereafter KM81) have also measured the integrated \hg\ flux, but this line is intrinsically fainter than \hb\ by more than a factor of two, and is more susceptible to interstellar extinction.

Nowadays, global fluxes in the red H$\alpha$ line are becoming the preferred benchmark, especially since the majority of PNe discovered over the last decade are mostly faint and reddened, and often undetected at \hb.  The bulk of these discoveries came from the MASH catalogues (Parker et al. 2006; Miszalski et al. 2008), which utilized the SuperCOSMOS \ha\ Survey (SHS; Parker et al. 2005).  We also note the new objects uncovered from the INT Photometric H-Alpha Survey (IPHAS; Drew et al. 2005), totalling 950 objects (Viironen et al. 2009a,b; Sabin et al. 2010, 2012;  L. Sabin, 2012, pers. comm.).   

To date, there have been considerably fewer \ha\ fluxes published in the literature. Some notable early efforts included Peimbert \& Torres-Peimbert (1971), Peimbert (1973) and Torres-Peimbert \& Peimbert (1977, 1979), while about 200 integrated  \ha\ fluxes for PNe were contributed by the Illinois group (e.g. Kaler 1981, 1983a,b; Kaler \& Lutz 1985; Shaw \& Kaler 1989, hereafter SK89), excluding fluxes derived from earlier photographic estimates (see Cahn \& Kaler 1971).  Thirty compact southern PNe have accurate integrated H$\alpha$, H$\beta$, H$\gamma$, \OIII, \NII\ and \HeII\ $\lambda$4686 fluxes determined by KM81, and 39 more PNe have accurate fluxes in several lines measured by Dopita \& Hua (1997, hereafter DH97).  Recent works include Ruffle et al. (2004) who determined diameters, \ha\ fluxes and extinctions for 70 PNe, and the accurate multi-wavelength data for six northern PNe published by Wright, Corradi \& Perinotto (2005, hereafter WCP05).  Other \ha\ flux measurements for smaller numbers of PNe are widely scattered through the literature.  

\subsection{Motivation for the Catalogue}\label{sec:motivation} 

Given the significant numbers of Galactic PNe that are currently being unearthed from recent narrow-band optical and infrared (IR) wide-field surveys,  the time is right to address the lack of systematic fluxes in the literature. This can now be done properly for the first time thanks to the availability of accurately calibrated emission-line surveys.  Tellingly, some of the brightest PNe in the sky have very few flux determinations.  The integrated \ha\ flux for the Helix nebula (NGC~7293) given here is only the second published in the literature, after Reynolds et al. (2005), while the \hb\ flux has been determined only twice (O'Dell 1962, 1998).  The Dumbbell nebula (M~27, NGC~6853) has been similarly neglected, with only two measurements of its integrated \hb\ flux (Osterbrock \& Stockhausen 1961; O'Dell 1998), and a single observation of its integrated \ha\ flux, an old one by Gebel (1968).  Furthermore, the solar neighbourhood (Frew \& Parker 2006) is dominated by the demographically common low-surface brightness (LSB) PNe,  typified by the discoveries of Abell (1966).  For these faint nebulae,  the H$\beta$ or H$\alpha$ fluxes are often poorly determined, if known at all.  The data published to date have been measured from a variety of techniques and are often inconsistent (cf. Kaler 1983b; Ishida \& Weinberger 1987; Kaler, Shaw \& Kwitter 1990; Pottasch 1996; Xilouris et al. 1996, hereafter XPPT), so it is obvious that more work needs to be done. 

Here we directly address these issues by presenting a homogenous catalogue of integrated \ha\ fluxes for 1258 Galactic PNe.  We incorporate some \ha\ fluxes derived from SHASSA and VTSS that were previously published by our group. These are for PFP~1 (Pierce et al. 2004), RCW~24, RCW~69 and CVMP~1 (Frew, Parker \& Russeil 2006), K~1-6 (Frew et al. 2011), M~2-29 (Miszalski et al. 2011),  Abell~23, Abell~51, and Hf~2-2 (Boji\v{c}i\'c et al. 2011b), and for the \HII\ region around PHL~932 (Frew et al. 2010).  

This paper is organised as follows:  In \S\,\ref{sec:surveys}, we outline the SHASSA and VTSS surveys used to provide the \ha\ fluxes, in \S\,\ref{sec:methods} we  describe the photometry pipeline and provide a discussion of the flux uncertainties, before describing our results, and the catalogue of fluxes,  in \S\,\ref{sec:results}.  We provide new independent reddening determinations for $\sim$270~PNe in \S\,\ref{sec:reddenings} and we outline our suggestions for future work in \S\,\ref{sec:future}.  We summarise our conclusions in \S\,\ref{sec:summary}.   
In addition, in Appendix~\ref{appendix_recal}, we provide corrected \ha\ fluxes for 49 objects, many little studied, from Abell (1966), Gieseking, Hippelein \& Weinberger (1986, hereafter GHW), Hippelein \& Weinberger (1990, hereafter HW90), Xilouris et al. (1994), XPPT, and Ali et al. (1997), re-calibrated to our common zero-point; 24 of these are not included in either Table~3 or Table~4.   Finally, as a resource for the wider community,  we provide total \ha\ fluxes for 76 misclassified objects from the literature in Appendix~\ref{mimic_fluxes}.

\section{The \ha\ Survey Material}\label{sec:surveys} 

The increasing online availability of wide-field digital imaging surveys in the \ha\ line is changing our ability to undertake large-scale projects such as this (for a review of earlier \ha\ surveys, see Parker et al. 2005).  Here we use two narrow-band CCD surveys: the Southern H$\alpha$ Sky Survey Atlas\footnote{http://amundsen.swarthmore.edu/} (SHASSA; Gaustad et al. 2001, hereafter GMR01) and the Virginia Tech Spectral-line Survey\footnote{http://www.phys.vt.edu/halpha} (VTSS; Dennison, Simonetti \& Topasna 1998), to provide the base data for our new PN flux determinations.  Other surveys such as the Mt Stromlo Wide Field \ha\ Survey (Buxton, Bessell \& Watson 1998) and the Manchester Wide Field Survey (e.g. Boumis et al. 2001) were unavailable in digital form so were not utilized in this study.

\subsection{SHASSA}\label{sec:SHASSA}

SHASSA is a robotic wide-angle digital imaging survey covering 21\,000 sq. degrees of the southern and equatorial sky undertaken with the aim of detecting \ha\ emission from the warm ionized interstellar medium (WIM).   The survey consists of 2168 images covering 542 fields south of +16\arcdeg\ declination, between Galactic longitudes of 195\arcdeg and 45\arcdeg\ at the mid-plane (GMR01).  SHASSA used a 52-mm focal length Canon lens operated at $f$/1.6, placed  in front of a 1024 $\times$ 1024 pixel Texas Instruments chip with 12$\mu$m pixels.  This produced images with a field of view of 13\arcdeg\ $\times$ 13\arcdeg\ (1014 $\times$ 998 pixels) and a scale of 47.64\arcsec\ per pixel.  The \ha\ interference filter ($\lambda_{\rm eff}$ = 6563\AA, FWHM = 32\AA) and a dual-band notch filter (that transmits two bands at $\lambda_{\rm eff}$ = 6440\AA\ and 6770\AA, both with FWHM = 61\AA) are mounted in a filter wheel in front of the camera lens (GMR01). 


There are four images available for each field: \ha, red continuum, continuum-corrected \ha\ (generated by subtracting each continuum image from the corresponding \ha\ image), and a smoothed \ha\ image.  The continuum-subtracted \ha\ images have a limiting sensitivity\footnote{The sensitivity limits of the various \ha\ surveys are generally quoted in Rayleighs, where 1\,R = 10$^{6}$/4$\pi$ photons cm$^{-2}$ s$^{-1}$ sr$^{-1}$ = 2.41 $\times$ 10$^{-7}$ erg cm$^{-2}$ s$^{-1}$ sr$^{-1}$ = 5.66 $\times$ 10$^{-18}$ erg cm$^{-2}$ s$^{-1}$ arcsec$^{-2}$ at H$\alpha$.} of better than 2\,R\,pixel$^{-1}$, corresponding to an emission measure of $\sim$4~cm$^{-6}$ pc.  The continuum-subtracted SHASSA data are available as either the original 48\arcsec\ resolution data, or smoothed data median-filtered to 5~pixels (4.0\arcmin),  which allows the detection of large-scale features as faint as 0.5\,R (or an emission measure of $\approx$1\,pc cm$^{-6}$).   GMR01 repeated all fields with offsets of 5\arcdeg\ in both coordinates, which helped confirm objects of very low surface brightness.  Thus, the number of individual measurements for a PN can be up to five in some favourable cases.

The full-resolution, continuum-subtracted SHASSA data often show pixels with unphysical, negative values which are residuals from poorly subtracted stellar images. These are largely removed in the smoothed images.  However these were not utilised for determining flux measurements, as smoothing mingles the flux from the PN with the ambient sky background, and can lead to an overestimate of the true flux when this background is significant.  

Despite the relatively coarse resolution of the SHASSA data, a large fraction of Galactic PNe are either large enough or bright enough to be readily detected by the survey, which allows for an accurate \ha\ flux determination.  The CCD camera  had a linear response up to the full-well capacity ($\approx$60\,000 ADU), but the variance ($\sigma^2$) increased non-linearly above 20\,000 ADU.  Note that almost all of the observed PNe give values below this limit.  Further details on the camera  and the data processing pipeline are given by GMR01. 

The SHASSA intensity calibration was derived using the PN spectrophotometric standards of DH97, after the continuum images had been scaled and subtracted from the \ha\ frames.  However, a difficulty in applying PN line fluxes to \ha\ narrow-band imaging is the proximity of the two \NII\ $\lambda\lambda$6548,~6584 lines which are included in the flanks of the SHASSA \ha\ filter bandpass. These vary in strength relative to \ha\ between PNe and may significantly affect the flux determination if not taken into account, especially for the so-called Type I nebulae (Peimbert 1978; Peimbert \& Serrano 1980; Peimbert \& Torres-Peimbert 1983; Kingsburgh \& Barlow 1994), which have elevated \NII/\ha\ ratios.  The DH97 standards were all compact so the \NII/\ha\ ratios are reliable. However, this ratio can vary quite significantly across well resolved PN and will be affected if the slit just happens to be oriented to intersect low-ionisation structures such as ansae and FLIERS (Balick et al. 1994; Gon\c{c}alves, Corradi \& Mampaso 2001).  We further discuss this point in \S\,\ref{sec:spectrophotometric}, below.
 
Calculating the transmission properties of the interference filter to these lines is also complicated by the blue-shifting of the bandpass with incident angle of the converging beam (e.g. Parker \& Bland-Hawthorn 1998).  These effects are considered in \S4 of GMR01 and are carefully accounted for in their calibration.  An additional uncertainty in the SHASSA zero-point derives from the contribution of geocoronal \ha\ emission.  GMR01 estimated this by comparing the SHASSA intensities with overlapping 1\arcdeg\ field-of-view data points from the Wisconsin H-Alpha Mapper (WHAM; Haffner et al. 2003), and interpolating if there is no available WHAM data.  WHAM provides a stable intensity zero-point over 70\% of the sky at 1\arcdeg\ resolution, and is now being extended to the southern sky by the WHAM-South Survey (Haffner et al. 2010).  Finkbeiner (2003, hereafter F03) showed there is no significant offset between WHAM and SHASSA data (cf. VTSS, see below), which indicates that the geocoronal contribution to the SHASSA \ha\ images has been appropriately corrected for.  

In order to further ascertain the reliability of the SHASSA intensity calibration, a first-step analysis was conducted by Pierce et al. (2004; see also Parker et al. 2005) which showed that the aperture photometry from the full-resolution data returns the best measurement of the integrated H$\alpha$ flux.  Available spectroscopic data were used to deconvolve the contribution from the [N\,{\sc ii}] lines passed by the SHASSA filter (see \S\,\ref{sec:spectrophotometric} and \ref{sec:pipeline} below) for these calibration PNe.  



\subsection{VTSS}\label{sec:VTSS}
A complementary survey to SHASSA, the VTSS (Dennison et al. 1998) covers a wide strip around the northern Galactic plane (15\arcdeg $<l<$ 230\arcdeg,  $|b|$ $\leq$ 30\arcdeg), north of $\delta$ $>$ $-15$\arcdeg.   Like SHASSA, the combination of fast optics, narrowband  interference filters, and a CCD detector gives this survey very deep sensitivity to diffuse \ha\ emission.  The VTSS used the Spectral Line Imaging Camera which utilised a Noct-Nikkor lens (f/1.2) of 58mm focal length placed  in front of a Tektronix 512 $\times$ 512 pixel chip with 27$\mu$m pixels.  This produced images with a circular field of view with a diameter of 10\arcdeg, with a resolution of 96\arcsec\ pixels.   A filter wheel in front of the lens holds a narrow bandpass \ha\ interference filter ($\lambda_{\rm eff}$ = 6570\AA, FWHM = 17.5\AA), various wider continuum filters, and an \SII\ filter centred at $\lambda$6725\AA.  Further details are given by Topasna (1999).

Each VTSS survey field was planned to have four images available: H$\alpha$,  continuum-corrected H$\alpha$, \SII\ and continuum-corrected \SII.  The continuum-corrected H$\alpha$ and \SII\ images are produced by subtracting an aligned and scaled continuum image from the H-alpha or \SII\ image.  Continuum images are taken with a wide-bandpass filter, or a double-bandpass filter astride the \ha\ line or \SII\ doublet (Dennison et al. 1998).  Each field name consists of the standard 3-letter IAU constellation abbreviation, plus a 2-digit running number (e.g. Cyg10).  However, VTSS  remains incomplete and is likely to remain so.  The original survey footprint was planned to cover $>$1000 sq. degrees (Dennison et al. 1998), but at the time of writing, continuum-corrected \ha\ images are available for only 106 of 227 fields (an additional six fields have uncorrected \ha\ images), and no \SII\ images are available.   The narrow bandpass of the VTSS \ha\ filter,  with essentially no transmission of the flanking \NII\ lines, makes the \ha\ flux determination straightforward in principle.  The sensitivity limit of VTSS is $\sim$1~R, comparable to SHASSA for diffuse emission, but since the resolution is only half that of SHASSA, the VTSS has a factor-of-four brighter detection limit for PNe (and other compact \ha\ emitters) due to confusion noise.

\section{Methodology}\label{sec:methods}

The following sections describe our flux measurement process, including the derivation of the input catalogue, the source references for the \NII\ fluxes which need to be deconvolved from the SHASSA fluxes, and detailed descriptions of the photometry pipeline and our error budget. 

\subsection{Input Catalogue}\label{sec:input_data}

The fluxes presented here are based on the contents of a `living' relational database we have designed (Boji\v{c}i\'c et al. 2013, in preparation) containing all bona fide, possible, and misclassified PNe from the major catalogues that have been published previously.  The PNe have been taken from Perek \& Kohoutek (1967), Acker et al. (1992, 1996), Kimeswenger (2001), Kohoutek (2001), the MASH catalogues (Parker et al. 2006; Miszalski et al. 2008), and the preliminary IPHAS discovery lists (Mampaso et al. 2006; Sabin 2008; Viironen et al. 2009a,b, 2011; Sabin et al. 2010; Corradi et al. 2011).  

To supplement these catalogues, we also added a sample of $\sim$320 true and candidate PNe from the recent literature, primarily discovered (or confirmed) optically.  These were taken from Zanin et al. (1997), Weinberger et al. (1999), Beer \& Vaughan (1999), Kerber et al. (2000), Cappellaro et al. (2001), Eracleous et al. (2002), Whiting, Hau \& Irwin (2002), Bond, Pollacco \& Webbink (2003), Boumis et al. (2003, 2006), Lanning \& Meakes (2004), Jacoby \& Van de Steene (2004), Su\'arez et al. (2006), Frew, Madsen \& Parker (2006), Whiting et al. (2007),  Miszalski et al. (2009), Fesen \& Milisavljevic (2010), Jacoby et al. (2010), Miranda et al. (2010), Acker et al. (2012), Parker et al. (2012b), Parthasarathy et al. (2012), and Kronberger et al. (2012).  

We also included over 100 unpublished candidates found by the Macquarie group,\footnote{So far 30 nebulae have been {\it discovered} from the SHASSA and VTSS; the nomenclature for those objects listed in this paper  continues on from Frew et al. (2006) and Frew (2008).} 
  plus sundry other objects, e.g. Mu~1 (A. Murrell, pers. comm. 2004) and Ju~1 (Jurasevich 2009).  All objects were cross-checked for duplication and identity, noting that the classifications of many nebulae have been fluid over time. An example is the Type~I planetary K~1-9 (PK\,219+01.1; Kondratyeva \& Denissyuk 2003) as noted in the SIMBAD database.

Additional candidate PNe recently found in near/mid-IR surveys were also added to our database.  These are unlikely to be detected in the SHASSA or VTSS surveys, but were included as we wanted to calculate the fraction of PNe found in the survey footprint that were detected by our photometry pipeline.  These PNe were taken from Phillips \& Ramos-Larios (2008), Kwok et al. (2008), Froebrich et al. (2011), Oliveira et al. (2011), Ramos-Larios et al. (2012) and Parker et al. (2012b), while another 320+ PN candidates found in the mid-IR at 24\,$\mu$m were added from Mizuno et al. (2010), Wachter et al. (2010) and Gvaramadze et al. (2010).

Over 6000 individual Galactic objects are currently in our working database, including 3320 bona fide and possible PNe\footnote{This total includes about 45 transitional objects (e.g. Su\'arez et al. 2006;  Parthasarathy et al. 2012).} and 480 post-AGB stars and related objects (Szczerba et al. 2007, 2012; Sahai et al. 2007; Lagadec et al. 2011).  There are also $\sim$1500 mimics of various kinds (e.g. Kohoutek 2001; Frew \& Parker 2010, hereafter FP10; Frew \& Parker 2011; Boissay et al. 2012),\footnote{Previous PN catalogues contain many astrophysical sources masquerading as PNe  (see Appendix~\ref{mimic_fluxes}).  We are currently reinvestigating the mimics in our database using a suite of multi-wavelength discrimination techniques (FP10; Cohen et al. 2011; Boissay et al. 2012).} with the remaining objects being currently unclassified; these await a more detailed investigation into their nature.  

Of the true and possible PNe, 2880 are located in the combined footprint of the two \ha\ surveys, which represent our input catalogue for the photometry pipeline (see \S\ref{sec:pipeline}).  This is a substantial increase in numbers over the preliminary input list presented by Boji\v{c}i\'c, Frew \& Parker (2012a).  In all, we measured \ha\ fluxes (see \S\,\ref{sec:pipeline}) for $\sim$1120 PNe from SHASSA and 178 PNe from the VTSS, 1234 objects in total because of overlap between the two surveys.\footnote{A few shock-excited pre-PNe such as CRL~618 were also detected by the pipeline.}    Thus $\sim$43\% of the PNe found in the SHASSA or VTSS surveys were bright enough to be detected by the pipeline.  For the other 1650 PNe, they were in the main too faint, or were northern PNe which fell outside the VTSS coverage (e.g. NGC~2392).  There were also a few moderately bright PNe which were too close to a bright star or artefact to be measurable.  We note that $\sim$80\% of MASH PNe are too faint to be detected in SHASSA, and almost all of the recent IPHAS discoveries are too faint for VTSS.  We have also measured the integrated \ha\ fluxes for 76 mimics and present these in Appendix~\ref{mimic_fluxes}.  If there is currently no consensus in the literature on the nature of a PN-like object, we included it in the main tables for the time being, and flagged it as a possible or uncertain PN.

\subsection{Spectrophotometric Data}\label{sec:spectrophotometric}

The VTSS \ha\ filter is essentially monochromatic, rejecting the flanking \NII\ lines (Topasna 1999) which simplifies the data reduction.  However, the SHASSA filter, while centred near \ha, is broader with a FWHM of 32\AA,  transmitting both \NII\ lines in the filter wings.  The transmission factors for the $\lambda$6548, \ha\ and $\lambda$6584 lines at rest wavelengths are 39\%, 78\% and 26\% respectively (GMR01).  In order to derive a pure H$\alpha$ flux for each object, the observed nebular $\lambda\lambda$6548,84/H$\alpha$ ratio (hereafter $R_{\rm \NII}$) is required to deconvolve the \NII\ contribution to the SHASSA red flux.  The spectrophotometric data required to do this is widely scattered in the literature, but we made a concerted effort to recover and evaluate these data.

We gave preference to any integrated values for $R_{\rm \NII}$ measured with modern linear detectors, or where we believe the data has been carefully taken and reduced. The primary sources for $R_{\rm \NII}$ are as follows: 
(i) large-aperture narrowband photoelectric photometry (e.g. KM81; SK89), (ii) narrowband CCD imagery (e.g. Hua, Dopita \& Martinis 1998),  (iii) wide-slit spectroscopy (e.g. Torres-Peimbert \& Peimbert 1977; Guti\'errez-Moreno, Cortes \& Moreno 1985;  DH97; WCP05; Wang \& Liu 2007), (iv) drift-scan long-slit methods (e.g. Liu et al. 2000, 2004; Tsamis et al. 2003; Zhang et al. 2005; Fang \& Liu 2011), (v) Integral Field Unit (IFU) spectroscopy (e.g. Tsamis et al. 2008), or (vi), data obtained with large-aperture Fabry-Perot spectrometers such as WHAM (Haffner et al. 2003) or other similar instrumentation (e.g. Lame \& Pogge 1996).  The WHAM data will be published in detail in a separate paper (G. Madsen et al., in preparation), but some preliminary flux data were presented by Madsen et al. (2006).  For these sources, we adopt a conservative uncertainty of 10\% in the value of $R_{\rm \NII}$.

For PNe without integrated values for $R_{\rm \NII}$, we adopted an average of the data presented in the {\it Catalog of Relative Emission Line Intensities Observed in Planetary Nebulae} (ELCAT)\footnote{http://stsdas.stsci.edu/elcat/} compiled by Kaler, Shaw \& Browning (1997),  supplemented with data taken from more recent papers in the literature, including Kraan-Korteweg et al. (1996), Pollacco \& Bell (1997), Weinberger, Kerber \& Gr\"obner (1997), Kerber et al. (1997, 1998, 2000), Perinotto \& Corradi (1998), Condon, Kaplan \& Terzian (1999), Ali (1999), Escudero \& Costa (2001), Rodr\'iguez, Corradi \& Mampaso  (2001), Milingo et al. (2002), Kondratyeva \& Denissyuk (2003), Kwitter, Henry \& Milingo (2003), Costa, Uchida \& Maciel (2004), Emprechtinger et al. (2004), Exter, Barlow \& Walton (2004), Costa, Uchida \& Maciel (2004), G\'orny et al. (2004, 2009), Krabbe \& Copetti (2006), Henry et al. (2010), Milingo et al. (2010), Miszalski et al. (2012a,b), Garc\'ia-Rojas et al. (2012) and Frew et al. (2012).  An average of measurements from several independent sources should be fairly representative of the integrated \NII/H$\alpha$ ratio for each PN, and we estimate a typical uncertainties in $R_{\rm \NII}$ of 10--30\%.  For the MASH PNe, as well as a significant number of other southern PNe, we utilised our extensive database of $>$2000 spectra taken as part of the MASH survey.  The analysis of these spectra will be published separately.  Long slit spectra obtained from the {\it San Pedro M\'artir} (SPM) {\it Kinematic Catalogue of Galactic Planetary Nebulae}\footnote{http://kincatpn.astrosen.unam.mx} (L\'opez et al. 2012) were also used to estimate $R_{\rm \NII}$ for objects with no other published data.  

However, it is possible that the value of $R_{\rm \NII} $ taken from long-slit spectra is systematically overestimated for some evolved PNe, as spectrograph slits are often placed on bright rims to maximise the S/N ratio for these low-surface brightness objects.  These brighter rims usually result from an interaction of the PN with the interstellar medium,  which are expected to have enhanced \NII\ emission (Tweedy \& Kwitter 1994, 1996;  Pierce et al. 2004).  If bright ansae or other low-excitation microstructures (Gon\c{c}alves et al.  2001) fall within the slit, this also has the potential to upset the \NII/\ha\ ratio.  Alternatively, the slit might only cover the central region of a PN where the excitation is higher, leading to an underestimate of $R_{\NII}$.  The derived  \ha\ flux might therefore be slightly in error in these cases.  Fortunately, the integrated \ha\ flux is only weakly sensitive to the exact value of $R_{\NII}$, for $R_{\NII} $ $<$ 1 (see equation~\ref{eq:nii_uncertainty}).
In the future, increasing numbers of PNe will be observed with the current and next generation of IFUs (e.g. Roth et al. 2004; Monreal-Ibero et al. 2005; Tsamis et al. 2008; Sandin et al. 2008), which will largely remove these problems.

\subsection{Photometry pipeline}\label{sec:pipeline}

To measure the PN fluxes, the IRAF\footnote{IRAF is distributed by the National Optical Astronomy Observatory, which is operated by the Association of Universities for Research in Astronomy (AURA) under cooperative agreement with the National Science Foundation.} {\tt phot} task was used.  A photometry pipeline was scripted in PyRAF, a Python interface to IRAF, by one of us (ISB) to facilitate the semi-automatic flux measurement of the large number of PNe found in the SHASSA and VTSS pixel data.  In the case of SHASSA many PNe fall in several (up to 5) separate fields so it is possible to obtain multiple independent flux measurements in these cases.  The amount of overlap between VTSS fields is much less, so the majority of PNe have only a flux measurement from a single field.

A PN was assumed to be detected if: (i) at least one pixel within the aperture had a flux of +5$\sigma$ above the adjacent sky background, and (ii), if rule (i) applies in more than 50\% of fields containing the PN. For each object, a circular aperture was automatically placed over the catalogued PN position, taken principally from the SIMBAD database (Wenger et al. 2007).   Circular apertures were chosen as most PNe are $<$2\arcmin\ (or $\sim$3 pixels) across, so any asymmetries are largely washed out. 
We refine the catalogued position using the {\it centroid} algorithm but with a maximum allowed shift from the original position of one pixel.  The PNe with a computed centroid more than one pixel from the catalogued position, and all larger PNe in our catalogue ($\theta_{PN}>3$\arcmin), were carefully examined and, if needed, we applied an appropriate correction to the aperture position.  For PNe that were detected on the images, but at less than 5$\sigma$ above the sky, we individually examined each object and if possible, manually measured an integrated flux.  These objects are flagged in the tables accordingly.

\begin{figure}
\begin{center}
\includegraphics[clip=true,scale=0.44]{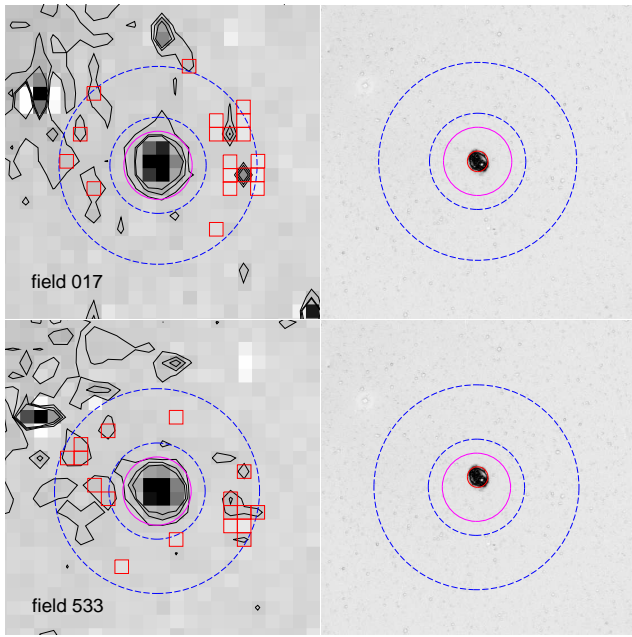}
\caption{An example showing the image cutouts for the evolved PN NGC~4071 (SHASSA left, SHS right) generated by the photometry pipeline. The top field is extracted from SHASSA field \#071 and the bottom is field \#533, and all images are 8\arcmin\ on a side with north-east at top left.   The left images plot the aperture diameter (magenta circle) and sky annulus (blue dashed circles), plus \ha\ intensity contours at 1$\sigma_{\rm sky}$, 3$\sigma_{\rm sky}$ and 5$\sigma_{\rm sky}$.  The right panels also indicate the catalogued PN diameter (small red circle).  Note the artefacts from the imperfect off-band continuum subtraction on the two SHASSA panels, which also show the marked-out pixels (red boxes) that are excluded by the photometry routine. A colour version of this figure is available in the online journal.
}
\label{fig:N4071}
\end{center}
\end{figure}

The radius in pixels of the aperture ($\theta_{ap}$) to be used was calculated from the estimated FWHM of the point spread function (PSF) and the angular size (major axis) of the nebula ($\theta_{PN}$) in arcsec, using:

\begin{equation}
\label{eq:apsize}
\theta_{ap} =
\begin{cases}
{\rm PSF} & \text{if } \theta_{PN} < {\rm PSF}\\
\\
{\rm PSF} + 0.25\,(\theta_{PN} - P_{S})/P_{S} & \text{if } \theta_{PN} \ge {\rm PSF}\\
\end{cases}
\end{equation}

where $P_{S}$ is the CCD plate scale (47.64\arcsec/pix for SHASSA and 96.4\arcsec/pix for  VTSS).
We chose~2.25 pix and 2.0~pix as the default PSF values, i.e. the minimum aperture radii, for SHASSA and VTSS respectively.   We found that an increase in the minimum aperture radius of 20\% affects the flux estimate by less than 5\%, however it significantly increased the chance of contamination of the measured flux from any nearby objects such as HII regions, or residuals from badly subtracted bright stars.  Hence, we adopted a smaller value for VTSS owing to its larger pixel size which increased the chance of contamination by nearby sources on the sky.  The uncertainty arising from the choice of the minimum aperture size ($\sigma_{\rm aper}$) was estimated by examining curves of growth between 2 and 3 pixels.   The applied routine automatically accounts for the differing relative areas of the object aperture and sky-subtraction annulus and it scales the sky subtraction accordingly. The number of pixels in a sky annulus is set to three times the number of pixels in the aperture (Merline \& Howell 1995). 


For each PN we also created a cutout image from the respective SHASSA or VTSS images, overlaid with circles indicating the catalogued PN diameter, the aperture and annulus sizes, and \ha\ intensity contours at 1$\sigma_{\rm sky}$, 3$\sigma_{\rm sky}$ and 5$\sigma_{\rm sky}$ above the measured sky level, where $\sigma_{\rm sky}$  is the standard deviation of pixel intensities in the sky annulus.  For PNe imaged by the SHS (Parker et al. 2005) including all of the MASH PNe (Parker et al. 2006; Miszalski et al. 2008), quotient images (H$\alpha$/ broadband $R$) were also generated for each PN.  The fits images were automatically created using the IRAF task {\tt imarith} before  being converted to {\tt png} format using the APLpy module\footnote{The Astronomical Plotting Library in Python, available from http://aplpy.github.com}).  Each image was linearly stretched using the default minimum and maximum pixel values (0.25\% and 99.75\%) and overlaid with the intensity contours from the corresponding SHASSA/VTSS image.  

This procedure allowed for rapid examination of the SHASSA and VTSS images (and SHS images where available) in common image viewing software for quality control.  The images were independently examined by all three co-authors and each object was noted as either a definite, possible, or non-detection on the SHASSA or VTSS images.  Any PNe that are confused with nearby objects or stellar residuals were flagged accordingly.

Finally, the total pixel counts, the flux after the background correction, and the statistics in the annulus and in the aperture were obtained for each PN.  Measurements affected by poor sky estimates, confusion with bright sources, or various stellar residuals or other artefacts contaminating the aperture or sky annulus were carefully examined and re-measured manually.  In these cases measurements of the sky background were made through an aperture identical to the PN aperture at a number of representative regions immediately surrounding the nebula, in order to accurately account for the surrounding diffuse \ha\ emission. The dispersion in these is the principal uncertainty in the measurement.

Many compact high surface brightness PNe have surrounding AGB halos (e.g. Corradi et al. 2003; Frew, Boji\v{c}i\'c \& Parker 2012a) included in the photometry aperture, but as the typical halo surface brightness is a factor of $10^{-3}$ fainter than the main shell (Corradi et al. 2003; Sandin et al. 2008; Sandin, Roth \& Sch\"onberner 2010), we ignore this contribution.  For NGC 7293, a structured AGB halo (Malin 1982; O'Dell 1998; Parker et al. 2001; Speck et al. 2002; O'Dell, McCullough \& Meixner 2004; Meaburn et al. 2005) is readily visible on SHASSA.  For others, such as NGC 3242 (Bond 1981; Phillips et al. 2009) and Abell~21, the surrounding material is more likely to be ionized ISM. These PNe are flagged in the main tables.

Obtaining integrated fluxes from the SHASSA data is quite straightforward.   Firstly, a `red' \ha+\NII\ flux in cgs units\footnote{1\,\ergcms\ = 1\,mW\,m$^{-2}$ in the $SI$ system.} is given by:

\begin{equation}
\label{eq:shassa_red}
F_{\rm red}^{\rm SHASSA} = 5.66\times10^{-18} \times P_{S}^{2} \times {\rm{\sc (SUM/10)}}  \quad   {\rm erg}\,{\rm cm}^{-2}\,{\rm s}^{-1}
\end{equation}

The constant in the expression is the conversion factor from Rayleighs to cgs units (at \ha\ and/or \NII) and as before, $P_S$ is the plate scale of the SHASSA survey (47.64\arcsec/pixel).  Note that the native SHASSA units are decirayleighs (1\,dR = 0.1\,R), hence the SUM of the background-corrected counts obtained from the routine are divided by 10.



As described in \S\,\ref{sec:SHASSA},  the SHASSA filter passes \ha\ and both \NII\ lines.   In order to derive a pure H$\alpha$ flux for each object, we used the observed \NII/\ha\ ratio (obtained as described in \S\,\ref{sec:spectrophotometric}) to deconvolve the \NII\ contribution to the SHASSA red flux.   The expression for correcting the \NII\ contribution is:

\begin{equation}
\label{eq:shassa_deconvolve}
F({\rm H}\alpha)_{\rm SHASSA} = \frac{F_{\rm red}}{K_{\rm tr}\,R_{\rm \NII} + 1}
\end{equation}

where $R_{\NII}$ is the adopted \NII/\ha\ ratio for the PN and $K_{\rm tr}$ is a dimensionless constant which takes into account the transmission of the filter at \ha\ and for each \NII\ line.  Note that the \NII\ flux refers to the sum of the $\lambda$6548 and $\lambda$6584 lines, and the ratio of the two is quantum mechanically fixed to be $\sim$3.  For the SHASSA filter we determine  $K_{\rm tr}$ = 0.375 using the transmission coefficients of the filter given by GMR01.  If only the brighter $\lambda$6584 line is measurable in a spectrum, $F$\NII\ is estimated as 4/3 $\times$ $F$($\lambda$6584).   


A smaller number of PNe were measured from the VTSS survey (Dennison et al. 1998).   The procedure for VTSS (including the treatment of the uncertainties) is identical to that used for the SHASSA measurements with the important simplification that a correction for \NII\ emission is not necessary as these lines are not passed by the narrowband VTSS \ha\ filter, unless the radial velocity of the PN is significant.    The native units of the VTSS are Rayleighs, so the adopted equation is:

\begin{equation}
\label{eq:vtss_flux}
F({\rm H}\alpha)_{\rm VTSS} = 5.66\times10^{-18} \times \,P_{S}^{2} \times \,{\rm SUM}  \quad   {\rm erg}\,{\rm cm}^{-2}\,{\rm s}^{-1}
\end{equation}

where in this case $P_S$ = 96.4\arcsec pix$^{-1}$.  The VTSS fluxes were then brightened by  0.04 dex (or 12\%) based on a comparison between SHASSA and VTSS fluxes (see \S\,\ref{sec:flux_comparison}).  

\subsection{Analysis of Photometric Errors}\label{sec:error_analysis}

The measured nebular flux for any PN  is expected to have a Poissonian uncertainty.  The total uncertainty in each flux measurement is a quadratic summation of the uncertainties due to photon statistics (i.e. shot noise from the nebula and sky), dark current noise, readout noise, the survey calibration uncertainty, and the uncertainty in the adopted [NII]/\ha\ ratio of the PN.  Additional terms are due to the PN's heliocentric radial velocity and the angular offset of the PN from the camera's optical axis (which causes a shift in the effective wavelength of the filter).   

A further potential uncertainty depends on any error in the angular position of the PN, which leads to vignetting of the Ôphotometric apertureÕ automatically placed around the catalogued PN position.  GMR01 found that the instrumental positions of several stars on each field matched the catalogued positions to 0.18 pixel (8\arcsec) rms, derived from first-order plate equations.  However, we found that a non-trivial fraction of Galactic PNe have uncertainties of up to 20\arcsec\ in the positions catalogued by SIMBAD.  For a small number of PNe with more severe positional errors, we manually re-centered the aperture to accept all of the flux from the PN.


The uncertainty due to shot noise scales as the square-root of the flux.  Following Masci (2008), the resulting uncertainty in the integrated red flux due to all noise terms,  including the uncertainty from the fixed minimum aperture ($\sigma_{\rm aper}$),  is given as:

\begin{equation}
\label{eq:phot_noise}
\sigma^2_{\rm phot}=  \frac{1}{g} \sum^{N_{A}}_{i} \frac{(S_{i}-\bar{B})}{N_{i}} + \left( N_{A} + k\frac{N_{A}^{2}}{N_{B}} \right) \sigma^{2}_{B/pix} + \sigma^{2}_{\rm aper}
\end{equation}

where $S_{i}$ is the recorded intensity, $g$ is the gain in electrons/ADU, $N_A$ is the number of pixels in the nebular aperture, $N_B$ is the number of pixels in the sky annulus, $N_i$ is the depth-of-coverage at pixel $i$ (for SHASSA, each survey field is the combination of five individual exposures so $N_i$ = 5), $S_i$ is the signal in pixel $i$, $\bar{B}$ is the mean sky background count per pixel in the annulus, and $\sigma^{2}_{B/pix}$ is the variance in the sky background annulus in [image units]$^2$/pixel.   We use an intensity-weighted mean, or centroid of the background pixel histogram in the sky annulus to use as the correction for the aperture; hence $k$ = 1 in the formula above (following Masci 2008).   Bad pixels or pixels with an intensity value $>$3$\sigma_{\rm sky}$ from the mean were rejected from the sky pixel distribution.

Fluxes were automatically flagged if the PN is confused with adjacent objects and stellar residuals (based on a variance cut in pixel values), or shows unphysical pixel values within the aperture.  While SHASSA is linear up to the full-well capacity (60\,000 ADU), the variance increases non-linearly above 20\,000 ADU (GMR01).  Consequently, 25 bright PNe on SHASSA were  flagged because the maximum pixel value was $>$20\,000 ADU.  Two PNe (IC~418 and NGC 6572) had a maximum pixel value slightly above 60\,000 ADU.  Any additional uncertainty added to the \ha\ flux, however, must be small, as our derived fluxes are in good agreement with published values.

An additional source of uncertainty follows from the shift in effective wavelength of the filter if the incident flux is not normal to its surface.  Because the \ha\ filter was placed in front of the lens, light entered the filter as a parallel beam at an angle of incidence equal to the angular distance of the object from the optical axis. For a simple interference filter, and at angles $\leq$10\arcdeg\ (Parker \& Bland-Hawthorn 1998), the wavelength,  $\lambda_\Theta$ at angle of incidence, $\Theta$ is given by :

\begin{equation}
\label{eq:transmission}
    \lambda_\Theta = \lambda_0 \left( 1-\frac{\sin^2\Theta}{{n_e}^2} \right)^{1/2} 
\end{equation}

where $\lambda_0$ is the central wavelength of the filter bandpass and $n_e$ is the refractive index of the spacer layers (usually $\sim$2).   As the angle of incidence increases, the \ha\ and [NII] lines shift toward the red with respect to the filter's central wavelength, altering the transmission coefficients of the filter (GMR01).  For $\Theta$ $\simeq$ 9\arcdeg, near the SHASSA image corners, the wavelength shift at \ha\ is approximately 10\,\AA, significant compared to the filter FWHM of 32\,\AA.  GMR01 found that the intensity-weighted average transmission of the two \NII\ lines is a weakly varying function of angle of incidence but did not consider this in their calibration.   However, to mitigate this effect, we do not measure the flux for any PNe that are more than 6\arcdeg\ from a field centre, unless this is the only available measurement.   There were such 18 cases, and these are flagged in Table~\ref{tab:shassatable}.




To calibrate the SHASSA data in ADUs per unit \ha\ flux, GMR01 compared their measured \ha\ fluxes for 18 compact PNe to the accurate \ha\ fluxes of DH97. The fluxes were corrected for the two \NII\ emission lines adjacent to \ha.   By definition, the mean difference between the PN fluxes in the GMR01 calibrating sample and those from DH97 is zero.  From this comparison, the gain factor of the survey was determined:  $g$ = 6.8 $\pm$ 0.3 ADU R$^{-1}$pix$^{-1}$ at \ha.    The calibration uncertainty ($\sigma_{\rm cal}$) of 8\%  was attributed to uncorrected atmospheric extinction and the undersampling of the instrumental PSF by the 12$\mu$m pixels.   Including an independent comparison of the integrated \ha\ flux of the Rosette nebula with that from Celnik (1983), the overall calibration uncertainty was found to be better than $\pm$9\% (GMR01).

A further term is the uncertainty in the deconvolved \ha\ flux due to the uncertainty in the adopted  \NII/H$\alpha$ ratio.  This is defined as:

\begin{equation}\label{eq:nii_uncertainty}
\sigma_{R}  =  \frac{0.375\,R}{(0.375\,R + 1)} \times \frac{dR}{R}
\end{equation}

where $R$ and $dR$ are the adopted \NII/\ha\ ratio and its proportional uncertainty respectively.  The total uncertainty on the \ha\ flux is then given by the following expression:

\begin{equation}\label{eq:tot_uncertainty}
\sigma^2_{F {\rm (H\alpha),~SHASSA}} =  {\sigma}_{\rm phot}^{2} +  {\sigma}_{R}^{2} +  {\sigma}_{\rm cal}^2
\end{equation}

where  $\sigma_{\rm phot}$ is the proportional uncertainty in the photometry measurement,  $\sigma_{R}$ is the uncertainty in the deconvolved \ha\ flux due to the uncertainty in $R_{\rm \NII}$, and $\sigma_{\rm cal}$ is the calibration uncertainty of $\pm$8\%. 

As described in \S\,\ref{sec:pipeline}, the VTSS fluxes were corrected based on zero-point offset between SHASSA/WHAM and VTSS (GMR01; F03; see \S\,\ref{sec:VTSS_fluxes}), noting that the WHAM survey provides a stable zero-point over 70\% of the sky (F03).
After applying this offset, the adopted VTSS calibration uncertainty was assumed to be equivalent to SHASSA, or 10\%.  This has been added in quadrature to the measurement uncertainty determined from the photometry routine to determine the final uncertainty on the \ha\ flux, viz:

\begin{equation}\label{eq:tot_uncertainty_VTSS}
\sigma^2_{F {\rm (H\alpha),~VTSS}} = {\sigma}_{\rm phot}^{2} +  {\sigma}_{\rm cal}^2
\end{equation}

where as before $\sigma_{\rm phot}$ and $\sigma_{\rm cal}$ are the measurement and calibration uncertainties respectively.

\subsection{Radial Velocity Corrections}

For PNe with known heliocentric radial velocities (e.g. Durand, Acker \& Zijlstra 1998), we calculated the expected shift of the \ha\ line in order to determine the filter transmission at that wavelength.  The vast majority of disk PNe have heliocentric radial velocities, $v_{\rm hel}  \leq \pm$100 \kms\ but bulge PNe can have relatively high velocities (up to $\pm$270 \kms; Durand et al. 1998), which can shift the observed wavelength by up to 6\AA.  The original calibration of GMR01 made no correction for the observed RV of the calibrating PNe, so any uncertainty introduced by this simplification should be already accounted for by the calibration uncertainty term, $\sigma_{\rm cal}$.  Because the exact filter transmission of the \ha\ and flanking \NII\ lines is difficult to model as the velocity changes, we simply add quadratically a further uncertainty of 6 per cent for bulge PNe with $|v_{\rm hel}|$ =  100--200 \kms, and 15 per cent for bulge PNe with $|v_{\rm hel}| >$ 200 \kms.  

For VTSS the change in filter transmission with wavelength is more rapid, due to the narrow FWHM of the filter. For a PN with a velocity of $-150$\kms, the \ha\ line is shifted toward the blue edge of the filter, and the \NII$\lambda$6584 is potentially brought into the filter bandpass.  Fortunately most northern PNe have relatively sedate radial velocities (only 5\% of PN have $v_{\rm hel}$ $<-100$ \kms). Again we assume that the calibration term, $\sigma_{\rm cal}$, accounts for this additional uncertainty.



\subsection{The effect of {\rm \SII} emission in the SHASSA continuum filter}

It is important to note that the \SII\ $\lambda\lambda$6717, 6731\AA\ lines are transmitted by the blue wing of the 6770\AA\ filter with throughputs of $\sim$16 and $\sim$3 per cent respectively, increasing towards the field edges.  The median \SII/\ha\ ratio for PNe from the data presented in FP10 is $\sim$0.10, and while some PNe can have \SII/\ha\ ratios approaching unity (FP10), these are evolved PNe with the \SII\ $\lambda$6717/$\lambda$6731 ratio in the low-density limit so any \SII\ contamination is lessened.  Fortunately the uncertainty in the \ha\ flux due to \SII\ oversubtraction is at the $\sim$1--2 per cent level  for a typical PN, and can be safely neglected. We assume any residual uncertainty is incorporated in the zero-point error of the survey (see later, \S\,\ref{sec:error_analysis}).  However, for strongly shock-excited nebulae such as supernova remnants and Herbig-Haro objects, the fluxes derived from the SHASSA images will be less accurate, perhaps with an uncertainty of up to 25\% fro objects near the edge of the fields  (see GMR01).


\subsection{Contamination of the nebular \ha\ flux by ionized helium}
We have made no correction to the \ha\ flux for the nebular  \HeII\ 6-4 line at $\lambda$6560\AA, which typically has an intensity of about 14\% of the  \HeII\ 4-3 line at $\lambda$4686\AA\ (Brocklehurst 1971; Hummer \& Storey 1987).  Given the range of excitation seen in PNe, this corresponds to a $\lambda$6560/$\lambda$6563 ratio ranging between zero for very low-excitation (VLE) PNe (Sanduleak \& Stephenson 1972), up to $\sim$5\% for the highest excitation objects (e.g. Kaler 1981).  Given the average intensity of the $\lambda$4686 line seen in PNe (CKS92; Tylenda et al. 1994), the \ha\ flux for most PNe is overestimated by  \lessim2\%.  This correction ($\sim$0.01 dex) is smaller than the observational uncertainties.

\subsection{Stellar contamination of the \ha\ flux}\label{CSPN_contamination}

PNe exhibit a wide diversity of central star types, with both emission-line and absorption-line spectra represented (Smith \& Aller 1969; M\'endez 1991; Tylenda, Acker \& Stenholm 1993; Werner \& Herwig 2006; DePew et al. 2011; Weidmann \& Gamen 2011; Werner 2012; Frew \& Parker 2012; Todt et al. 2012; Miszalski et al. 2012b; Boji\v{c}i\'c et al. 2012b).  It is possible that PNe with bright central stars relative to the surrounding shell have integrated \ha\ fluxes slightly in error.  While the continuum subtraction process should largely alleviate this, there remains the possibility that contamination by strong emission-line central stars (e.g. [WC] stars) will lead to an overestimated flux for the PN.  We examined the spectra of several [WC] stars presented by DePew et al. (2011) and measured the strengths of the Pickering\,6 \HeII\ line  at $\lambda$6560\AA.  We found this line is typically at an intensity  level of \lessim2 per cent of the nebular \ha\ flux, so we made no change to the integrated fluxes as this correction ($\sim$0.01 dex) is also smaller than the observational uncertainties for most of the PNe considered.

Similarly, for PN with central binaries dominated by bright B-, A- or F-type companions with strong Balmer absorption lines, the nebular \ha\ flux may be underestimated.  The PNe affected are NGC~2346 (A5\,V), NGC~3132 (A2\,IV-V), Hen~2-36 (A2\,III), Abell~14 (B7\,V), Abell~79 (F0~V) and SuWt~2 (A1\,IV sb2); there is no available VTSS image for NGC~1514 (A0\,III).  We estimated a magnitude offset (\ha\ -- $R$) from the typical equivalent width of the \ha\ line for each spectral type (e.g. Jaschek \& Jaschek 1987), and then derived a correction factor to the \ha\ flux for each PN based on a comparison of the apparent stellar $R$-band magnitude with the integrated nebular magnitude.  The correction factors are relatively small, ranging from 0.04~dex (Abell~79) to 0.12~dex (SuWt~2), and have been applied to the \ha\ fluxes presented in Tables~\ref{tab:shassatable} and \ref{tab:vtss_table}.  

However, for luminous LBV, B[e], and WR stars in ejecta nebulae, the contribution of the emission line star to the total \ha\ flux is potentially significantly greater. For those objects listed in Table~B1, we have made no attempt to deconvolve the \ha\ (or \HeII\ $\lambda$6560) flux of the ionizing star from the total \ha\ flux, so the nebular fluxes should be considered as firm upper limits.


\section{Results}\label{sec:results}

Our new homogeneously measured \ha\ fluxes  for 1230 true and possible PNe, derived from both the SHASSA and VTSS data, are presented in the following sections.

\subsection{SHASSA \ha\ Fluxes}

The final \ha\ fluxes for each PN from $n$ separate SHASSA fields are combined to give a weighted mean:

\begin{equation}
\label{eq:weighted_mean}
    \bar{F_w} = \frac{ \sum_{i=1}^n w_i F_i}{\sum_{i=1}^n w_i}
\end{equation}

where [$F_{1}, F_{2} \ldots \,F_{n}$] are the individual flux estimates, with associated weights [$w_{1}, w_{2} \ldots \,w_{n}$] determined from the inverse variances, $w_{i} = 1/\sigma_{i}^{2}$, via equation~\ref{eq:tot_uncertainty}. 
It follows that the variance of the weighted mean is:

\begin{equation}
\label{eq:weighted_error}
\sigma^2_{\bar{F}}  =  \frac {V_1} {V_1^2-V_2} \sum_{i=1}^n w_i \left(F_i - \bar{F}_{w} \right)^2 
\end{equation}
   
where $V_1 = \sum_{i=1}^n w_i$  and $V_2 = \sum_{i=1}^n {w_i^2}$.
\smallskip


Table~\ref{tab:shassatable} contains the mean H$\alpha$ fluxes for $\sim$1120 PNe derived from SHASSA images.   Columns~1 and 2 give the standard $PN G$ designation and the common name respectively, and columns 3 and 4 give the right ascension and declination (epoch J2000.0).  The adopted \NII/H$\alpha$ ratio is given in column~5, and the logarithms of the averaged red flux and the corrected \ha\ flux are given in columns~6 and 7 respectively.  Column~8 lists the adopted aperture radius in arcmin, and column 9 gives the number of independent fields from which a measurement is obtained. The derived extinction is given in column~10 and any notes are indexed in column 11, including noting if the PN has a central star with Wolf-Rayet features (see \S\,\ref{CSPN_contamination}).   

If we do not have any information on the \NII/H$\alpha$ for an individual PN, we simply list the integrated red flux as measured from the images. An \ha\ flux can be derived using equation~\ref{eq:shassa_deconvolve}  once spectroscopic data becomes available in the future.

\begin{table*}
{\footnotesize	
\begin{center}
\caption{\ha\ fluxes for 1120 true and possible PNe measured from SHASSA.  The table is published in its entirety as an online supplement. A portion is shown here for guidance regarding its form and content.}
\label{tab:shassatable}
\begin{tabular}{clrrccrcccc}
\hline
PN G &  Name & RAJ2000 & DEJ2000 & $R_{\rm \NII}$ & logF$_{\rm red}$ & logF(H$\alpha$)& $r_{\rm aper}$    & ${\rm N}_f$& c$_{\beta}$ &Note\\
\hline
108.4$-$76.1&BoBn 1&00:37:16.0&$-$13:42:58&0.2&$-$12.03&$-$12.05$\pm$0.06&1.8&3&~0.00~~&C\\
118.8$-$74.7&NGC 246&00:47:03.4&$-$11:52:19&0.0&$-$10.07&$-$10.07$\pm$0.03&6.1&2&0.02&\\
255.3$-$59.6&Lo 1&02:56:58.4&$-$44:10:18&1.1&$-$10.99&$-$11.14$\pm$0.06&4.5&4&$\ldots$&\\
220.3$-$53.9&NGC 1360&03:33:14.6&$-$25:52:18&0.0&$-$9.78&$-$9.78$\pm$0.03&9.4&2&0.00&6\\
206.4$-$40.5&NGC 1535&04:14:15.8&$-$12:44:22&0.0&$-$9.95&$-$9.95$\pm$0.03&5.0&1&0.04&\\
$\ldots$&Fr 2-3&04:56:20.0&$-$28:07:48&1.6&$-$10.89&$-$11.09$\pm$0.12&9.5&1&$\ldots$&1\\ 
243.8$-$37.1&PRTM 1&05:03:01.7&$-$39:45:44&0.0&$-$12.02&$-$12.02$\pm$0.07&1.8&3&$\ldots$&5\\
215.5$-$30.8&Abell 7&05:03:07.5&$-$15:36:23&0.8&$-$10.30&$-$10.41$\pm$0.06&11.0&1&$\ldots$&\\
205.8$-$26.7&MaC 2-1&05:03:41.9&$-$06:10:03&0.0&$-$11.77&$-$11.77$\pm$0.05&1.8&3&$\ldots$&\\
190.3$-$17.7&J 320&05:05:34.3&10:42:23&0.0&$-$10.91&$-$10.92$\pm$0.04&1.8&3&0.04&\\
\hline
\end{tabular}
\end{center}
}
\begin{flushleft}
Notes:  (1) Possible PN; (2) pre-PN; (3) transition object; (4) uncertain counts; (5) confused with nearby object; (6) bad pixels in aperture; (7) object near field edge; (8) flux excludes halo; (9) flux corrected for CSPN;  (10) Wolf-Rayet CSPN; (N) previously unpublished object; (V) very low excitation PN; (C) specific comment given.
\end{flushleft}
\end{table*} 


\subsubsection{Comparison of fluxes from different SHASSA fields}\label{sec:duplicates}

\begin{figure}
\begin{center}
\includegraphics[clip=true,scale=0.53]{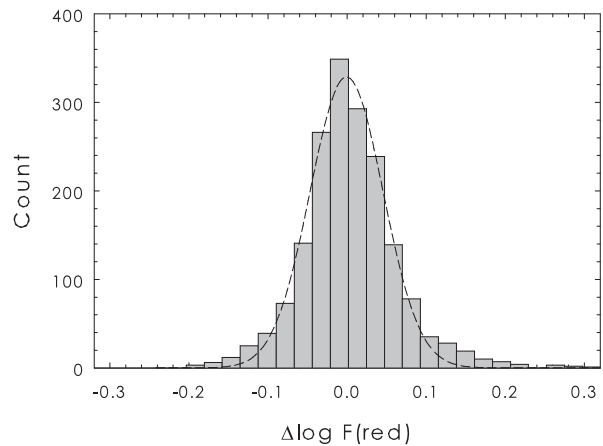}
\caption{A difference histogram of individual flux measurements from SHASSA.  The abscissa plots the difference between an individual flux measurement and the mean for that object ($\Delta$log\,$F$), and the ordinate shows the number of separate determinations.  The distribution is approximately Gaussian, with a standard deviation of 0.05 dex, in excellent agreement with the nominal zero-point error of the Survey.}
\label{fig:duplicates}
\end{center}
\end{figure}

For each PN detected on SHASSA, we compared the individual flux measurements from separate fields with the mean flux of each PN to ascertain the repeatability of our procedure.  In the difference histogram (Figure~\ref{fig:duplicates}) the abscissa plots the difference between an individual flux measurement and the mean for that object ($\Delta$log\,$F$), and the ordinate shows the number of determinations based on our measurements of all PNe detected in at least three SHASSA fields.  The distribution is approximately Gaussian, with a dispersion of 0.05 dex, in excellent agreement with the nominal zero-point error of the survey.

\begin{figure}
\begin{center}
\includegraphics[clip=true,scale=0.54]{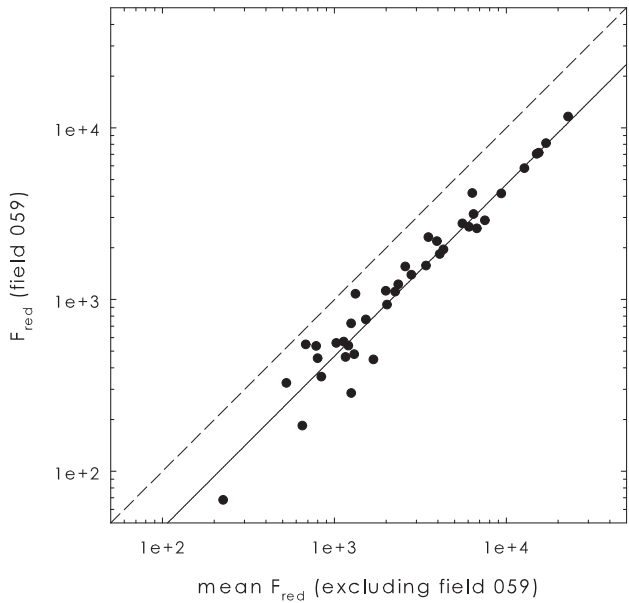}
\caption{A comparison of the \ha+\NII\ fluxes from SHASSA field \#059 with the remainder.  The abscissa plots the mean fluxes of 32 PNe from overlapping fields (excluding field \#059) and the ordinate plots the fluxes for these PNe measured from field \#059.  The dashed and dotted lines show the 1:1 relation and the least squares fit to the data respectively.  The fluxes from field \#059 are a factor of 2.14 too low (or offset by 0.33 dex).  }
\label{fig:059}
\end{center}
\end{figure}

During this process it was noticed that the integrated fluxes measured from SHASSA field \#059 were discrepant compared to the fluxes measured from overlapping fields.  This field has an erroneous zero-point calibration, and gives nebular \ha\ fluxes too faint by a factor of 2.14 (see Figure~\ref{fig:059}).  We found smaller offsets for eight other fields.  The correction factors for these fields are presented in Table~\ref{table:shassa_corrections}.  Five of the nine fields are below $\delta = -60$\arcdeg, where a cross-check with WHAM data is unavailable (GMR01).  The source of error for the four fields north of this limit is not clear.

From equation~\ref{eq:shassa_deconvolve} it follows that at $R_{\NII}$ = 2.7, the contribution from the two \NII\ lines in the overall flux is equal to \ha.  We use this value to divide our PN sample into two groups, \NII-dominated and \ha-dominated, and compare the SHASSA \ha\ fluxes with our own VTSS measurements and with independent literature values.  We found no significant difference between the samples, but did find that the \ha\ fluxes from the Illinois group tend to be overestimated for PNe with $R_{\NII}$ $\geq$\,2.  We attribute this to their \ha\ filter passing an uncorrected amount of \NII\ emission (Kaler, Pratap \& Kwitter 1987; SK89).




\begin{table}
{\footnotesize	
\begin{center}
\caption{Offsets applied to nine discrepant SHASSA fields.  The correction  factor was found by dividing the mean flux of all PNe in fields excluding the affected field by the mean flux of the same PNe in that field.}
\label{table:shassa_corrections}
\begin{tabular}{cccr}
\hline
SHASSA field 	&       Field centre 									& Corr. factor		& Offset 				\\
				&      (J2000) 										&					& (dex)  				\\
\hline
031				&$08^{\rm h}53^{\rm m}	-$60\arcdeg11\arcmin\		& 0.80				&	0.10~   			\\  
059 				&$16^{\rm h}04^{\rm m}	-$50\arcdeg08\arcmin\		& 2.14				&	$-$0.33~		\\  
172 				&$06^{\rm h}42^{\rm m}	-$10\arcdeg03\arcmin\		& 0.81				&	0.09~			\\  
190				&$18^{\rm h}43^{\rm m}	-$09\arcdeg57\arcmin\		& 1.34				&	$-$0.13~		\\ 
518 				&$14^{\rm h}49^{\rm m}	-$75\arcdeg12\arcmin\		& 1.17				&	$-$0.07~		\\  
536 				&$15^{\rm h}55^{\rm m}	-$65\arcdeg09\arcmin\		& 1.40				&	$-$0.15~		\\  
584				&$15^{\rm h}13^{\rm m}	-$45\arcdeg11\arcmin\		& 1.50				&	$-$0.18~		\\  
653 				&$18^{\rm h}36^{\rm m}	-$24\arcdeg57\arcmin\		& 1.33				&	$-$0.12~		\\  
708 				&$07^{\rm h}02^{\rm m}	-$05\arcdeg04\arcmin\		& 1.25				&	$-$0.10~		\\  
\hline
\end{tabular}
\end{center}
}
\end{table}

\subsubsection{Comparison of SHASSA and literature \ha\ fluxes}\label{sec:flux_comparison}

\begin{figure*}
\begin{center}
\includegraphics[clip=true,scale=0.55]{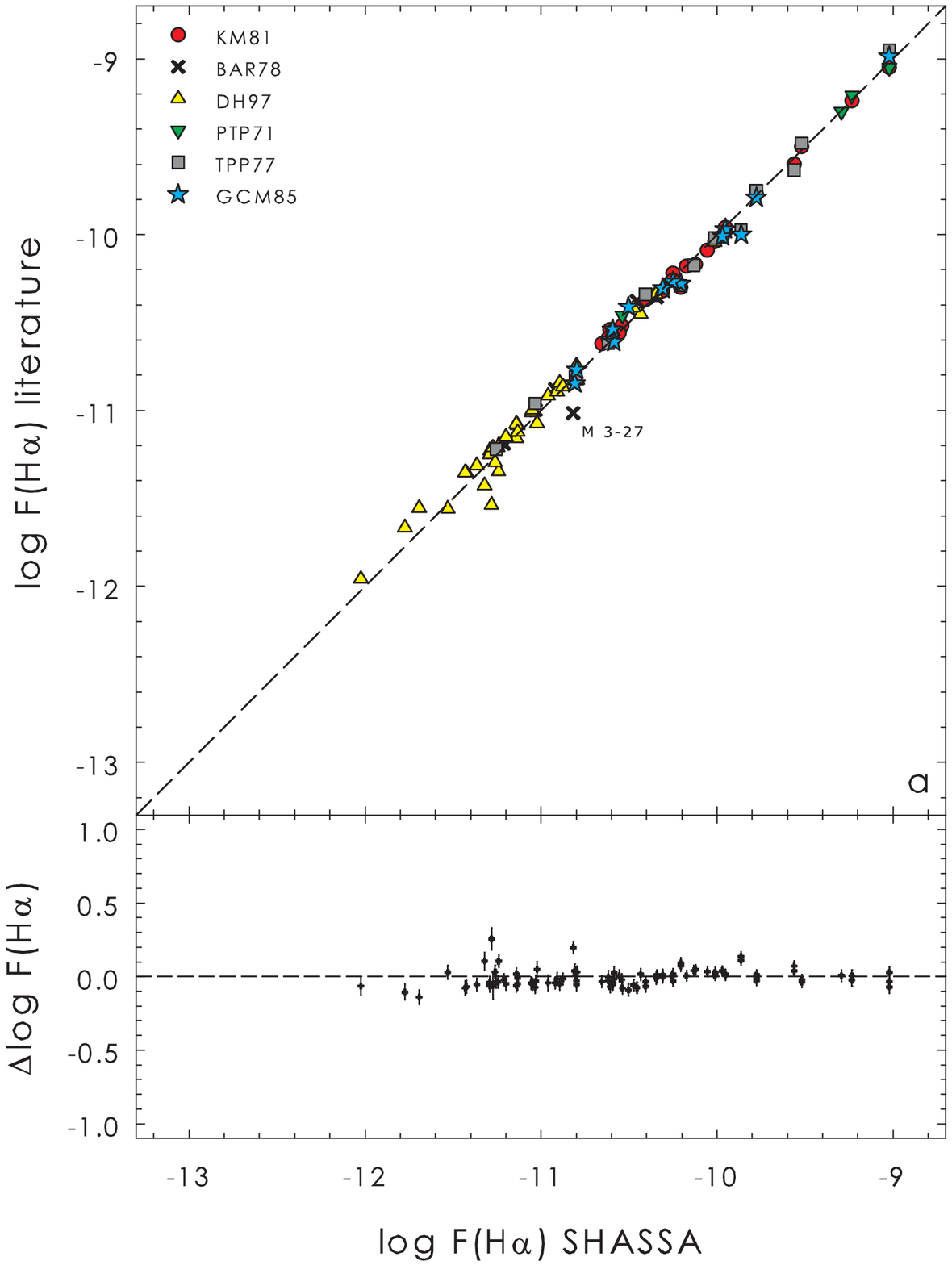}
\includegraphics[clip=true,scale=0.55]{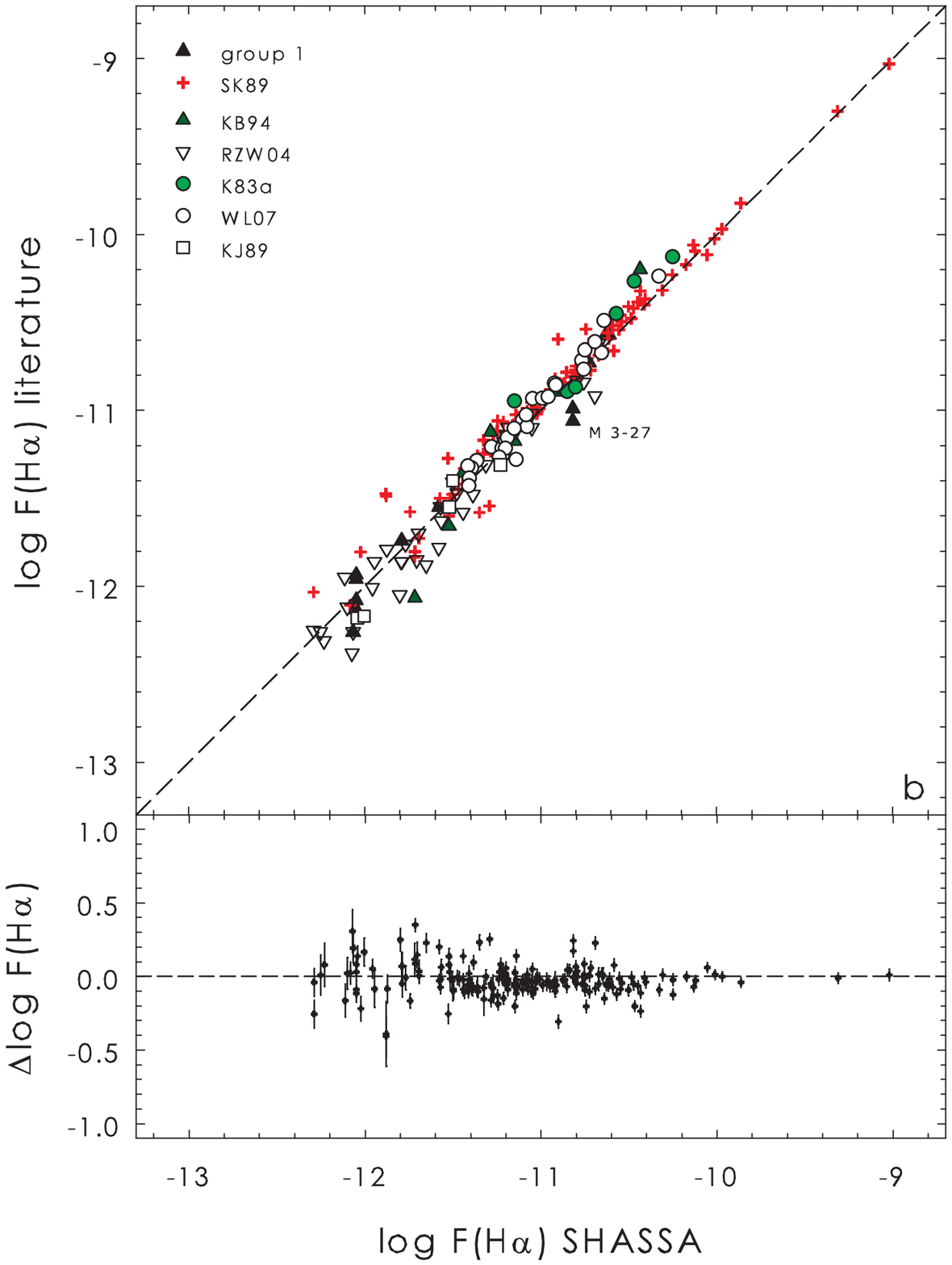}
\caption{A comparison of our \ha\ fluxes determined from SHASSA with those from the literature.  The left panel shows fluxes from sources deemed to have the highest level of precision and accuracy, as defined in the text.  The agreement between our fluxes and those from the literature is excellent over more than three orders of magnitude.  The right panel compares our \ha\ fluxes with the fluxes from sources with a somewhat lower accuracy, or reliable data-sets with less than three objects in common. A colour version of this figure is available in the online journal. }
\label{fig:comp1}
\end{center}
\end{figure*}

In this section we present a detailed comparison between our new SHASSA \ha\ fluxes and the equivalent measurements for the same PNe available in the literature. We demonstrate the validity of these new SHASSA fluxes and highlight some problems with previous determinations.  The early study of Pierce et al. (2004) showed that the deconvolved SHASSA \ha\ fluxes agreed with published data to $\Delta$\,F(H$\alpha$)~=~--0.00\,dex, $\sigma$ = 0.07\,dex in the sense of SHASSA minus literature fluxes (see also Parker et al. 2005).   Since GMR01 used 18 bright PNe from DH97 as SHASSA calibrators, it was expected that a comparison between the full set of DH97 fluxes and those derived here would have a negligible zero-point offset.  We indeed found no offset for the 18 PNe of the original calibrating sample, and $\Delta$\,F(H$\alpha$) = $-$0.02 dex for the full set of 36 objects, excluding PN~G307.2-09.0 (Hen 2-97) as it has an erroneous \ha\ flux (M. Dopita, pers. comm. 2012).   Not only does our expanded analysis here give a useful cross-check to the GMR01 intensity calibration, but it also allows the veracity of the adopted aperture photometry technique used here to be ascertained, including the treatment of the deconvolution of the [N\,{\sc ii}] lines from the red flux for each PN.    

In order to undertake the comparison, we have carefully compiled a large database of \ha\ and other line fluxes from the literature.  The literature fluxes have been measured using a variety of methods, with a range of observational uncertainties.   In several references (e.g. Barker 1978; Kaler 1980; Kaler \& Lutz 1985; SK89), we readily calculated the integrated \ha\ fluxes from the tabulated \hb\ fluxes and quoted \ha/\hb\ ratios.  However, in other references (Peimbert \& Torres-Peimbert 1971;  Torres-Peimbert \& Peimbert 1977; Guti\'errez-Moreno, Cortes \& Moreno 1985), only the observed \hb\ flux was given along with the reddening-corrected \ha/\hb\ ratio for each object.  We re-determined the observed \ha/\hb\ ratio, and hence the observed \ha\ flux, by using the logarithmic extinctions given for each PN and applying the reddening law used by the authors.   In some cases, only the \ha\ surface brightnesses were provided (Boumis 2003, 2006).  We calculated the \ha\ fluxes using the nebular diameters given in these papers.  In addition, some studies only provide global \ha+\NII\ fluxes (e.g. Xilouris et al. 1994, 1996).  The treatment of these data is described in Appendix~\ref{appendix_recal}.

A correction was then applied to the older \ha\ values due to the flux recalibration of Vega (Hayes \& Latham 1975; Oke \& Gunn 1983).  Those fluxes published before 1975 were decreased by 0.02\,dex  (0.03 dex for PTP71), while the fluxes tied to the fainter calibration of Miller \& Mathews (1972), e.g. Ahern (1978), were brightened by 0.07 dex  (see Shaw \& Kaler 1982; SK89, and CKS92 for further comments).

\begin{table}
{\footnotesize	
\begin{center}
\caption{Statistical comparison of our \ha\ fluxes measured from the SHASSA with independent values from the literature.  Only data-sets with more than three PNe in common are listed.
The table is organized into four tranches according to the contents of Figure~\ref{fig:comp1} and Figure~\ref{fig:comp2} (see text for details).}  \label{table:ha_comparison}
\begin{tabular}{lrccclc}
\hline
Reference	&   $\Delta$$F$	   & 	  $\sigma$ 			     &    $n$ 		&   Meth.  &	Det.	   		&  Fig.	\\
\hline
BAR78		&   $-$0.03	          & 		0.03			     	     &	5	       &	Spec   &    IDS	       &  	\ref{fig:comp1}a	        \\   
DH97		&   $-$0.02		   & 		0.08			            &	37	       &	Spec    &   CCD		 &  	\ref{fig:comp1}a		 \\  
GCM85		&   	  0.01	  	   & 		0.06		   		     &	13	       &	Spec   &    PEP        	&  	\ref{fig:comp1}a	        \\
KM81    		&       0.00	          & 		0.03			    	     &	25	       &	Aper   &   PEP	        &  	\ref{fig:comp1}a		 \\  
PTP71		&  $-$0.02		  & 		0.05			            &	4	       &	Spec   &   PEP 		&  	\ref{fig:comp1}a	        \\  
TPP77		&  $-$0.02 	          & 		0.09			            &	14	       &	Spec   &    IDS	         &  	\ref{fig:comp1}a	        \\  
\hline	
K83a   		&    	$-$0.09		  & 	       0.12				     &  	6	 	&	Aper    &    PEP	         &  	\ref{fig:comp1}b		\\  
KB94		&    $-$0.03		   & 	       0.19				     &  	12	 	&	Spec   &   IDS	         &  	\ref{fig:comp1}b		\\  
KJ89  		&   	0.06		  	  & 		0.11			            &	5	       &	Aper   &    CCD           &  	\ref{fig:comp1}b	        \\ %
RZW04		&   	0.03  		   & 		0.11		           	     &	43	       &	Aper   &    CCD        	 &  	\ref{fig:comp1}b	       \\
SK89		&    $-$0.05	          & 		0.10		  	            &	98	   	&	Aper   &   PEP	      	  &  	\ref{fig:comp1}b		\\  
WL07		&    	$-$0.06		  & 		0.08		  	            &	30	   	&	Spec   &   CCD	      	  &  	\ref{fig:comp1}b		\\  
\hline	
B01, B03 	&   	$-$0.25	   	   & 		0.29			            &	15	       &   Sp/Ap   & CCD   		&  	\ref{fig:comp2}a	       \\
BDF99	 	&     0.04	          & 		0.35		            	     &	11	       &	Aper   &     CCD         &  	\ref{fig:comp2}a	       \\ 
Bm03, Bm06	&   	0.26		          & 		0.28			            &	10	      &	Aper    &     CCD         &  	\ref{fig:comp2}a	       \\  
CSBT  		&   	$-$0.03  	   & 		0.48 	            	     &	5	      &	Aper   &      CCD        &  	\ref{fig:comp2}a	       \\ 
HDM98		&   	0.03  		   & 		0.16	           		     &	14	      &	Aper    &     CCD         &  	\ref{fig:comp2}a	       \\  
HK99 		&   	0.10	 		   & 		0.21		    	            &	4	       &	Aper   &      CCD        &  	\ref{fig:comp2}a	       \\  
K81     		&     0.06		   & 		0.13				     &	5		&	Aper   &   PEP	      	  &  \ref{fig:comp2}a		\\  
K83b		&     $-$0.02		   & 		0.14		   	            &      16	       &	Aper   &    PEP	     	  &  	\ref{fig:comp2}a		\\  
KSK90 		&   	0.06 		   & 		0.34			            &	26	        &	Spec &      IDS           &  	\ref{fig:comp2}a	       \\  
MFP06	     	&   	$-$0.12		   & 		0.12		           	     &	10	        &	F-P	  &     CCD            &  	\ref{fig:comp2}a	       \\  
RCM05		&   	0.01	  	   	   & 		0.11			            &	16	        &	F-P	  &     CCD           &  	\ref{fig:comp2}a	       \\  
\hline	
ASTR91A 	&     0.00	  	   & 		0.18			            &	207	        &	Spec  &      IDS         &  	\ref{fig:comp2}b	       \\   
ASTR91B 	&   	0.09  		   & 		0.26			            &	225        &	Spec  &       IDS       &  	\ref{fig:comp2}b	       \\	
ASTR91C	&   	0.21		   	   & 		0.44			            &	161	        &	Spec  &       IDS       &  	\ref{fig:comp2}b	       \\	
VV68		&   	$-$0.37		   & 		0.28			            &	105	   	 &	ObP	  &      pg            &  	\ref{fig:comp2}b	       \\  
VV75		&    $-$0.28	  	   & 		0.20			            &	16	   	&	ObP	  &      pg            &  	\ref{fig:comp2}b	       \\  
\hline		
\end{tabular}
\end{center}
}
{\scriptsize
Reference codes for Table~\ref{table:ha_comparison}, Table~\ref{table:ha_comparison_vtss}, Figure~\ref{fig:comp1} and Figure~\ref{fig:comp2}: 
\smallskip \\
ASTR91: Acker et al. (1991); B01: Bohigas (2001); B03: Bohigas (2003); BAR78: Barker (1978); BDF99: Beaulieu et al. (1999); Bm03: Boumis et al. (2003); Bm06: Boumis et al. (2006); CSBT: Cappellaro et al. (2001);  DH97: Dopita \& Hua (1997); GCM85: Guti\'errez-Moreno et al. (1985);  HDM98: Hua et al. (1998); K83a: Kaler (1983a); K83b: Kaler (1983b); KJ89: Kwitter \& Jacoby (1989); KM81: Kohoutek \& Martin (1981a); KSK90: Kaler et al. (1990); KB94: Kingsburgh \& Barlow (1994);  MFP06: Madsen et al. (2006); PTP71: Peimbert \& Torres-Peimbert (1971); RCM05: Reynolds et al. (2005); RZW04: Ruffle et al. (2004); SK89: Shaw \& Kaler (1989); TPP77: Torres-Peimbert et al. (1977); VV68: Vorontsov-Vel'yaminov et al. (1968); VV75: Vorontsov-Vel'yaminov et al. (1975); WL07: Wang \& Liu (2007). \\
}
\end{table}


For completeness we also provide the references which had three or less PNe in common with our catalogue (and for which we did not include any statistical information in the tables).  For plotting purposes, we group these sources into two groups, based on the relative accuracy of the data sets.  The first group includes \ha\ fluxes from Adams (1975), Ahern (1978), Cuisinier et al. (1996), Dufour (1984), Hippelein, Baessgen \& Grewing (1985),  Hawley \& Miller (1978a), Kaler (1980), Kaler \& Lutz (1985), Kelly et al. (1992), Mavromatakis et al. (2001b), Miszalski et al. (2011), Moreno et al. (1994), Peimbert et al. (1991), Pe\~na et al. (1990), Torres-Peimbert et al. (1981), WCP05, Wesson \& Liu (2004), Zhang \& Liu (2003), and Zijlstra et al. (2006).

The second group includes fluxes from Ali \& Pfleiderer (1997), Goldman et al. (2004), Hawley (1981), Hawley \& Miller (1978b), Hua (1988), Hua \& Grundseth (1985), Jacoby \& Van de Steene (2004), Kaler (1981), Kaler \& Hartkopf (1981), Kistiakowsky \& Helfand (1993), Kwitter et al. (2003), Lame \& Pogge (1996), Liu et al. (2006), Louise \& Hua (1984), Sahai et al. (1999), Pe\~na, Torres-Peimbert \& Ruiz (1991), Sahai, Nyman \& Wootten (2000), Shen, Liu \& Danziger (2004), Torres-Peimbert, Peimbert \& Pe\~na (1990), Turatto et al. (1990), Xilouris et al. (1994), and the corrected fluxes from Ali, Pfleiderer \& Saurer (1997) and Xilouris et al. (1996).  Also included here for plotting purposes are the fluxes from Bm03, Bm06, CSBT, and HK99 (see Table~\ref{table:ha_comparison} for details on these data sets). 


Table~\ref{table:ha_comparison} presents a detailed statistical comparison between our SHASSA \ha\ fluxes and those from the most important literature sources.
The columns give respectively the comparison dataset,  the mean $\Delta$log F(\ha) between data-sets (in the sense SHASSA $-$ literature fluxes), the standard deviation of the difference,  the number of PNe in common, and the technique and detector used for each data-set.  A positive value of the offset means our fluxes are brighter.  This table can be compared with Table~3 of Kovacevic et al. (2011), which details a comparison of the integrated \OIII\ fluxes for $\sim$170 PNe between various sources.

\begin{figure*}
\begin{center}
\includegraphics[clip=true,scale=0.55]{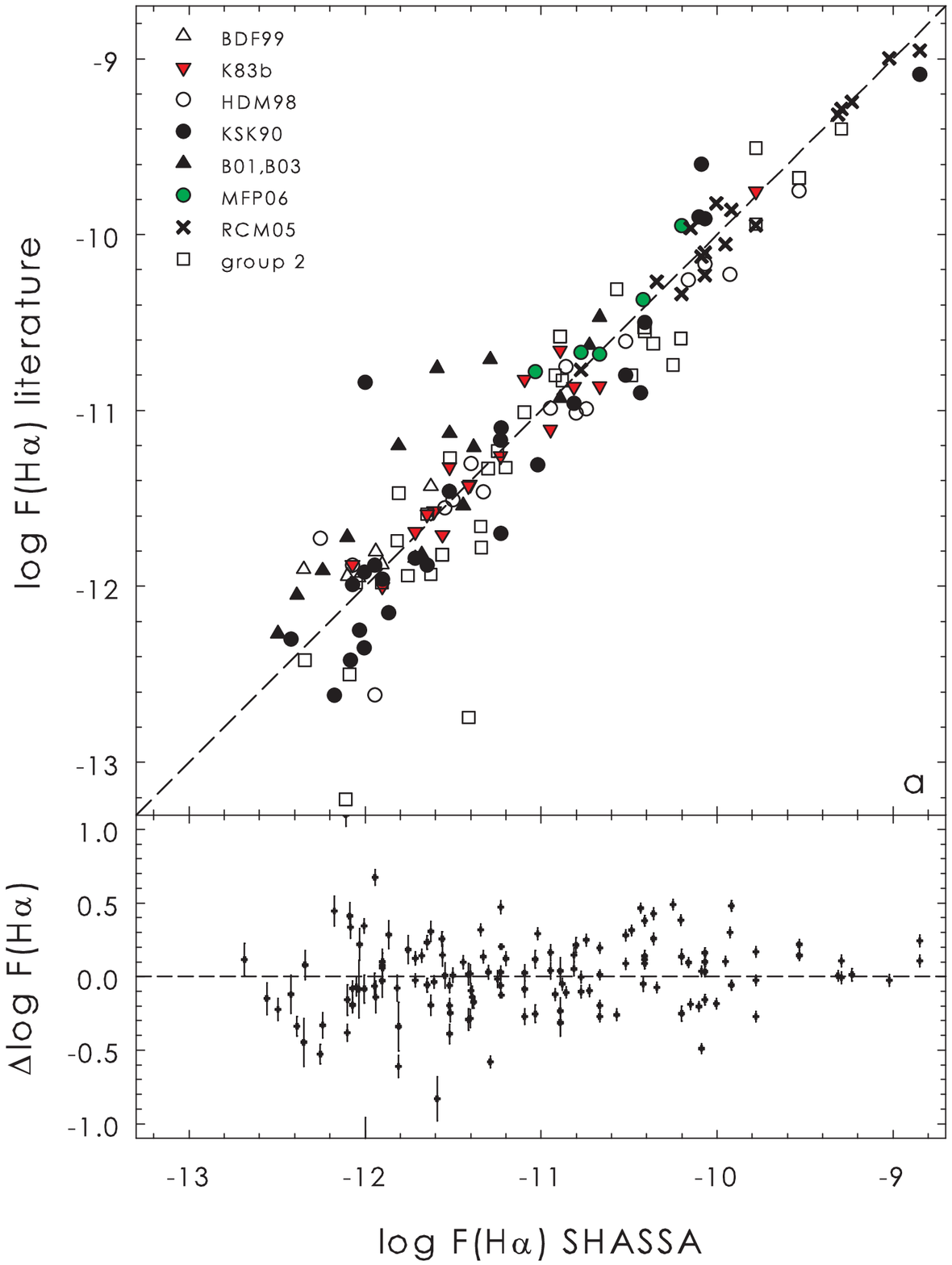}
\includegraphics[clip=true,scale=0.55]{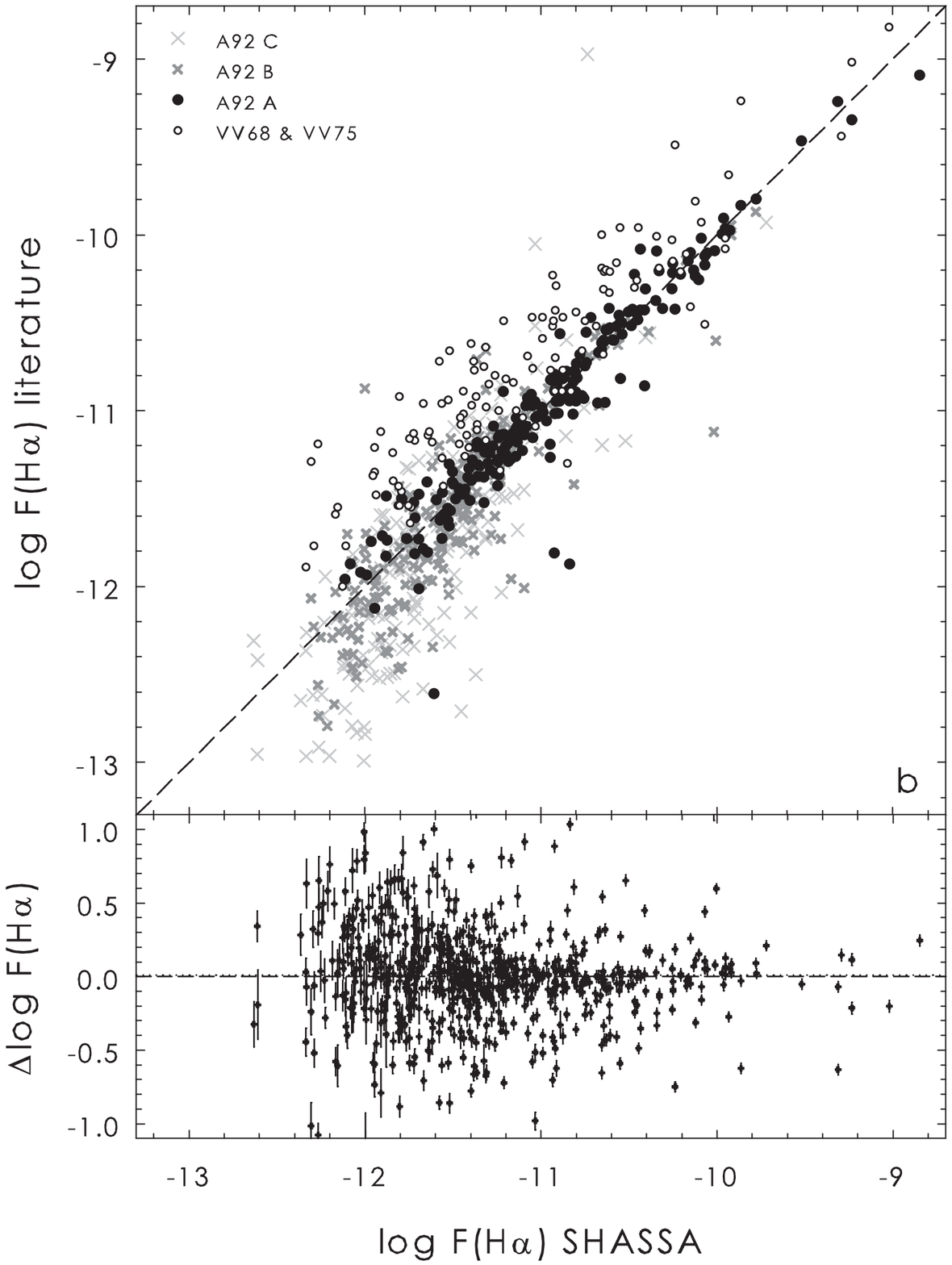}
\caption{The left panel compares our SHASSA with literature \ha\ fluxes for PNe of low surface brightness (see the text for details).  The right panel compares our \ha\ fluxes with the two largest data-sets from the literature (VV and ASTR91), as described in the text.  A colour version of this figure is available in the online journal.  }
\label{fig:comp2}
\end{center}
\end{figure*}

Figure~\ref{fig:comp1} and Figure~\ref{fig:comp2} graphically illustrate these comparisons, with each sample being identified with a different symbol.  These samples are separated according to our opinion on their reliability,  the PN's surface brightness, and the measurement techniques employed.  In Figure~\ref{fig:comp1}a (left panel) we plot our \ha\ fluxes (excluding those flagged as less certain) for 101 mostly compact PNe against fluxes from independent data sets that meet two quality criteria: (i) a mean flux difference from ours, $\Delta F$(\ha) $\leq$0.05~dex {\it and} (ii) a dispersion around this mean, $\sigma$ $\leq$ 0.10 dex. These data sets include the SHASSA calibrating measurements of DH97, the high-quality aperture fluxes of Kohoutek and Martin (1981), and other reliable \ha\ fluxes in a series of papers by the Peimberts (refer to Table~\ref{table:ha_comparison}). The lower panels of the figure give the difference between the data-sets, in the sense SHASSA minus literature fluxes.   The statistical results from these comparisons appear in the first tranche of entries in Table~\ref{table:ha_comparison}.   The \ha\ fluxes from KM81 show the best agreement with our data, with a negligible offset and a dispersion around the mean of only 0.03\,dex.




Other than the calibrating sample of GMR01 and our own work, we are not aware of any other published \ha\ fluxes for PNe using SHASSA data, with the exception of McCullough et al. (2001) who determined a flux for Abell~36 of $F$(\ha) = $2.8\times10^{-11}$ \ergcms, which is 28\% lower than our own determination of $3.9\times10^{-11}$ \ergcms.  We attribute this difference to the smaller  (4\farcm8 diameter) aperture used by McCullough et al. (2001), which missed some flux from this 6\farcm0 $\times$ 4\farcm7 planetary.

In Figure~\ref{fig:comp1}b (right panel), we plot the \ha\ fluxes of 150 PNe which we deem to be of slightly lower accuracy, based on the data-set meeting only one of the above criteria.  Some objects were excluded from the comparison, as the published measurement is not an integrated flux, e.g. NGC~5189 (SK89).  The statistical results appear as the second tranche of entries in Table~\ref{table:ha_comparison}.  The literature \ha\ fluxes have their own associated uncertainties (typically $\pm$0.01--0.05 dex), but based on the 252 fluxes presented in Figure~\ref{fig:comp1}, we find SHASSA to be calibrated to better than $\pm10$\% across nearly the whole survey.  This is in agreement with the nominal error supplied by GMR01, excluding those fields with an incorrect zero point (see Table~\ref{table:shassa_corrections}).


Of the brighter PNe (log$F$(\ha)$>-11.0$) plotted in Figure~\ref{fig:comp1}, the most discrepant point is M~3-27, an unusual object in several respects.  The three previous \ha\ fluxes in the literature that are deemed to be reliable (Adams 1975; Ahern 1978; Barker 1978) agree within their respective uncertainties,\footnote{The \hb\ flux of Kohoutek \& Martin (1981b) also agrees with the \hb\ fluxes from Adams (1975), Barker (1978) and Ahern (1978). The \ha\ flux of Kohoutek (1968) agrees with our own, but is derived from photographic plates and has a substantial uncertainty.} but our \ha\ measurement is about 0.2 dex brighter (Figure~\ref{fig:comp1}).  While our measurement is only from one SHASSA field (\#262),\footnote{The PN is at the very edge of field 261, and some flux is lost.} which has no zero-point offset, the detection is good and there is no confusion of the PN with any nearby sources.  This observation (epoch 1999.30)  is $\sim$25 years after the earlier measurements, suggesting a secular change in the flux with time.  This is a very compact, dense and presumably young object (Adams 1975; Miranda et al. 1997) and we initially thought that the change was due to the possible evolution of the central star, a situation seen in the very young object Hen~3-1357 (Parthasarathy et al. 1993; Bobrowsky 1994).  However, M 3-27 shows evidence for Balmer self-absorption (Adams 1975; Ahern 1978; Barker 1978), and the bulk of the \ha\ (and radio-continuum emission) originates in an ionized, optically-thick stellar wind which appears variable with time (Miranda et al. 1997).  M~3-27 also exhibits large variations in the measured electron densities and temperatures (Barker 1978; Feibelman 1985), similar characteristics to the well-known variable PN, IC 4997 (Aller \& Liller 1966; Hyung, Aller \& Feibelman 1994). Both these nebulae have very high \OIII$\lambda$4663/\hg\ ratios, overlapping with the symbiotic stars in the diagnostic plot of  Guti\'errez-Moreno, Moreno \& Cortes (1995).  Interestingly, the fluxes for M~3-27 presented by Wesson, Liu \& Barlow (2005) show that some line ratios have changed markedly between the mid-1970s and 2001, however, Wang \& Liu (2007) state that the ``fluxes for a number of lines published by WLB05 $\ldots$ seem to be incorrect.'' Unfortunately the latter authors did not  provide any updated data.   Notwithstanding this, our new flux adds weight to the conclusion of Miranda et al. (1997) that the \ha\ emission is variable.  This is certainly an interesting object, and warrants further study.

In Figure~\ref{fig:comp2}, we present two panels of lower quality flux comparisons. In Figure~\ref{fig:comp2}a (left panel), we compare our SHASSA fluxes for 138 mostly LSB PNe ($S_{\rm H\alpha}<10^{-4}$ \ergcms sr$^{-1}$) with the equivalent measurements from different literature sources.  The ability to recover fluxes for these faint PNe depends on an accurate background subtraction, given the often low data counts recorded above sky.  Such LSB PNe are generally quite extended and may have an asymmetric or irregular morphology, further complicating flux measurement.  The statistical summary of these comparisons appear in the third tranche of results in Table~\ref{table:ha_comparison}.  Unsurprisingly the mean offsets and dispersions are up to a factor of four larger for these samples than those of Figure~\ref{fig:comp1}, along with several large outliers.   
We excluded some literature fluxes from this comparison because the largest photometer aperture used was much smaller than the nebular diameter.  This leads to very uncertain values in the estimated total flux in these cases (e.g. Abell~21, Abell~35, Jones~1 and K~2-2; Kaler 1983b).   The WHAM \ha\ fluxes (Reynolds et al. 2005; Madsen et al. 2006) are measured with a 1\arcdeg\ beam, so contaminating background emission may occur for some PNe.  One example is Abell~74, where diffuse background emission is located near to the PN. This object is omitted from the comparison table.  We also omit the flux for Abell~60 from Hua et al. (1998), which is too faint.  The SHASSA \ha\ fluxes determined here are all integrated fluxes, measured through an aperture that was always larger than the nebula on the CCD image.  They are to be preferred for these large evolved PNe.

In Figure~\ref{fig:comp2}b (right panel), we present 760 lower quality fluxes from ASTR91 and Vorontsov-Vel'yaminov et al. (1968, 1975).  The comparison of our \ha\ fluxes with the older photographic determinations of Vorontsov-Vel'yaminov et al. (1968, 1975) was done primarily for historical interest.  These fluxes were derived from objective-prism plates which suffered from high non-linearity (e.g. Torres-Peimbert 2011).  The fluxes are not reliable and are systematically brighter than our data by 0.2 -- 0.4\,dex, increasing in offset at faint flux levels.   

ASTR91 determined integrated \hb\ fluxes for 880 PNe, with the majority scaled up from small-aperture slit spectroscopic measurements.  We calculated \ha\ fluxes from these data using the observed \ha/\hb\ ratios given by ASTR91.  These authors also compiled \hb\ fluxes for $\sim$350 PNe with accurate independent measurements from the literature.  We then calculated \ha\ fluxes for these objects using the \ha/\hb\ ratios from ASTR91.  These make up ASTR91 sample `A', with typical flux errors of $\leq$0.08\,dex.  This subsample gives the best agreement with our SHASSA  data, with a small offset and a reasonably low dispersion of 0.14\,dex.  For PNe that were larger than the slit aperture used (typically $\theta >$ 5\arcsec), ASTR91 determined the integrated \hb\ flux by scaling up the observed flux by a geometric ratio (ratio of the PN area to the aperture area), before applying an empirical correction factor.  Again we determine the integrated \ha\ flux for each PN using their published \ha/\hb\ ratio. These objects make up sample `B', including 224 PNe with an uncertainty $\leq$0.15 dex.    

For the larger PNe ($\theta >$ 1\arcmin), the slit covered only a small fraction of the object, so the derived integrated fluxes become very uncertain. These lower-quality fluxes (ASTR91, sample `C') have errors $>$0.15 dex, and are not reliable, as the extrapolation from the observed flux through a small aperture to the full dimensions of the PN introduces a large uncertainty.  Figure~\ref{fig:comp2}b shows that errors of 0.5 -- 1.0 dex are not uncommon, so we recommend that the data from ASTR91 B and C samples be only used if there is no other data source.  Their quoted uncertainties also appear to have been underestimated for these objects, a point previously noted by Ruffle et al. (2004) and WCP05.  A similar effect, though less severe, is seen in the data from Kaler et al. (1990) and Bohigas (2001, 2003).  These samples form the final tranche of entries in Table~\ref{table:ha_comparison}.

\begin{figure*}
\begin{center}
\includegraphics[clip=true,scale=0.925,angle=0]{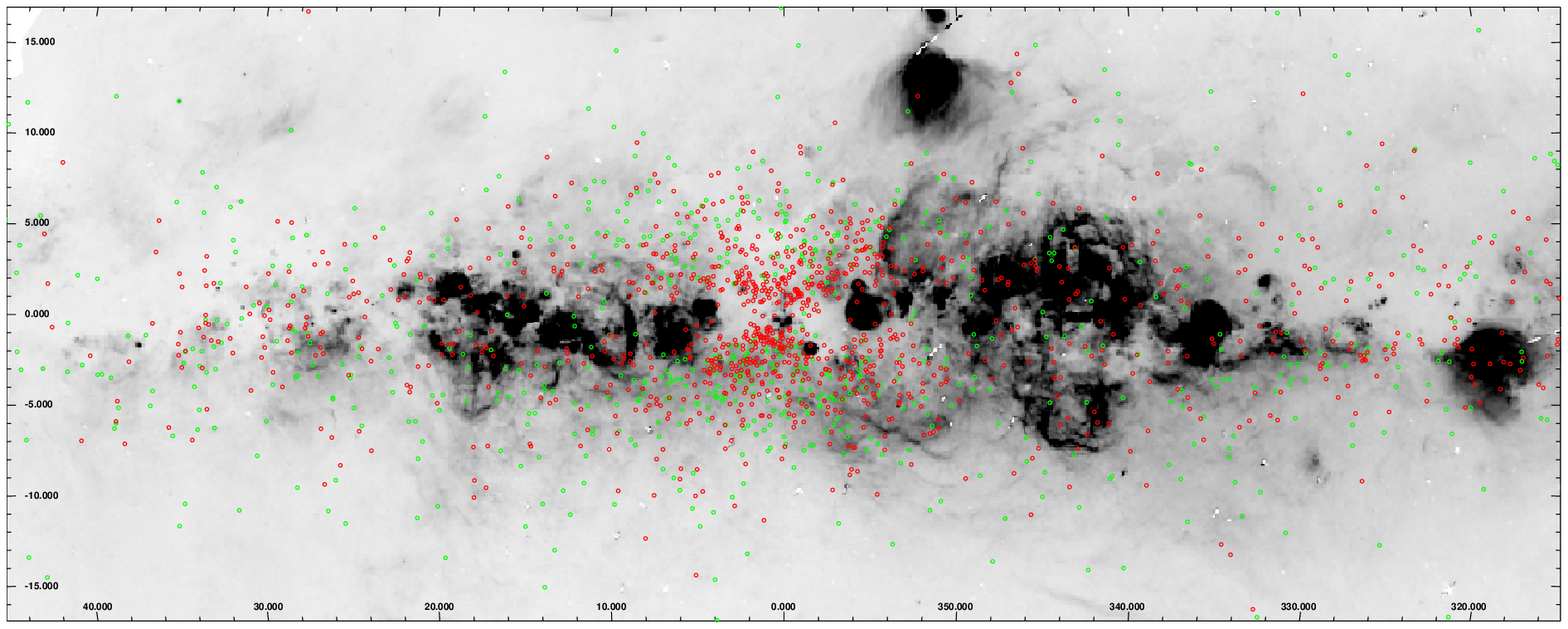}
\includegraphics[clip=true,scale=0.925,angle=0]{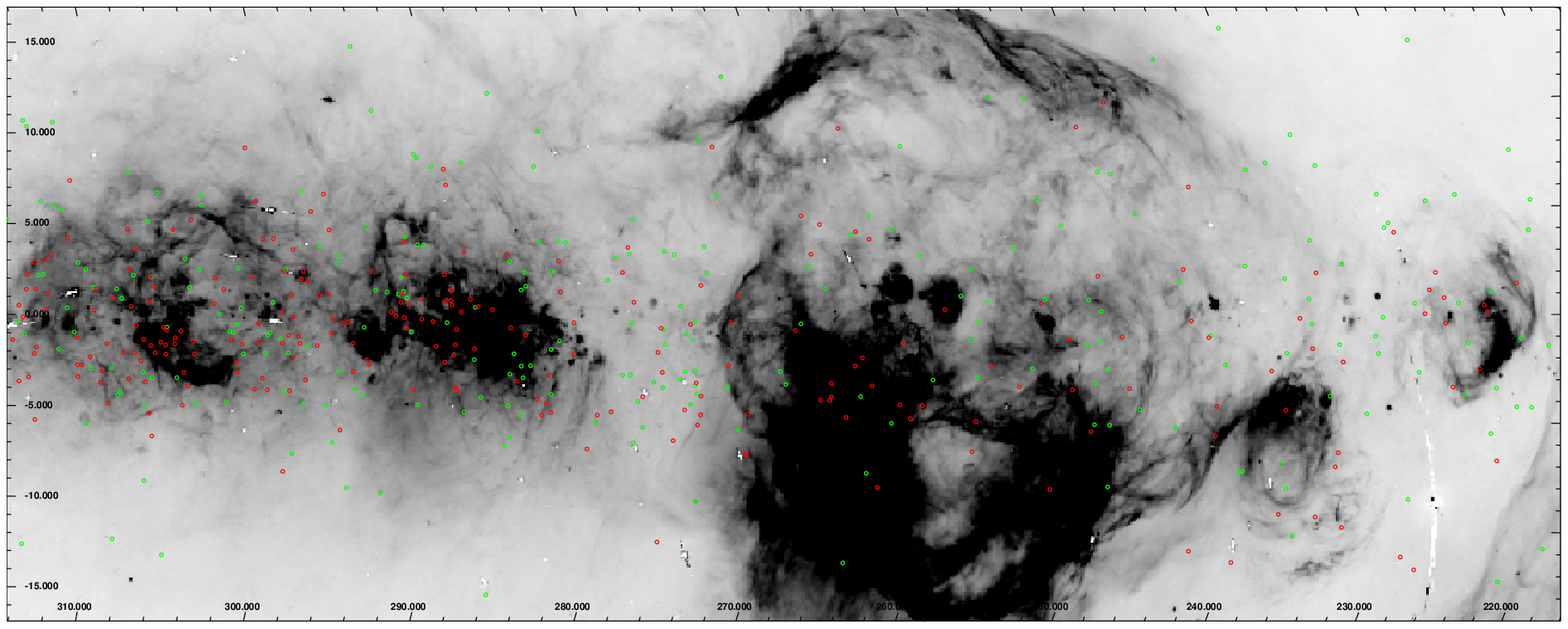}
\caption{The spatial distribution of PNe overlaid on a greyscale \ha\ map of the southern Galactic plane generated from SHASSA data.  The top panel is centred on the Galactic Bulge ($l$ = 0\arcdeg), while the lower panel is centered at $l$ = 270\arcdeg, in the vicinity of the Gum nebula.  The green circles represent SHASSA-detected PNe, while red circles mark non-detections.  As expected the latter are concentrated in high-extinction regions close to the mid-plane, in the crowded star fields of the Bulge, or in areas of bright background \ha\ emission. A colour version of this figure is available in the online journal.}
\label{fig:shassa_overlay}
\end{center}
\end{figure*}

Lastly, as SHASSA provides nearly 90 per cent of the new \ha\ fluxes in our catalogue, we show in  Figure~\ref{fig:shassa_overlay} the spatial distribution of PNe measured on SHASSA, plus non-detections over a greyscale \ha\ map of the southern Galactic plane.   The green circles represent detections, while the red circles are non-detections.  As expected the non-detections are mostly faint PNe, concentrated in regions of high extinction close to the mid-plane, in the crowded star fields of the Galactic bulge, or in areas of complex emission where the poor resolution prevents detection against the bright ambient background.   The SHASSA flux limit is log\,F(\ha) $\simeq$ $-12.8$, and this limit is $\sim$0.5--1~dex brighter for areas close to the Galactic plane with a high level of diffuse \ha\ emission.   As an example of an object near the  plane, we note the faint bipolar PHR~1315-6555 with log\,F(\ha) = $-12.45 \pm 0.01$ (Parker et al. 2011).  This PN resides in an area of  background \ha\ emission and is only barely detected on SHASSA at the 1$\sigma$ level.

\subsection{VTSS \ha\ Fluxes}\label{sec:VTSS_fluxes}

In a similar fashion to the SHASSA fluxes, we present  VTSS \ha\ fluxes for 178 northern PNe. The resolution of VTSS is coarser than SHASSA and owing to the increased level of confusion at low intensities, the VTSS flux limit is brighter; for regions with low uniform sky background, the VTSS limit is log\,F(\ha) $\simeq$ $-12.2$ compared to log\,F(\ha) $\simeq$ $-12.8$ for SHASSA.  
As noted previously, the median intensity in each VTSS image has been set equal to zero, so the survey can be considered to have an arbitrary zero point (Dennison et al. 1998).  From a comparison of two VTSS fields, Aql04 and Ori11, with the equivalent binned SHASSA data, GMR01 estimated correction factors of 0.10 and 0.07 dex respectively, which one needs to apply to the VTSS data in order to get the same surface brightness for diffuse emission measured in both surveys.  F03 applied this same factor to all VTSS images in order to compare the VTSS data with SHASSA, using available WHAM data as a cross-check, finding the resulting agreement between VTSS and SHASSA to be good (see his figure 7).  This offset factor was then assumed to be constant across the whole VTSS survey.

\begin{table*}
{\footnotesize	
\begin{center}
\caption{\ha\ fluxes for 178 true and possible PNe measured from VTSS.  The table is published in its entirety as an online supplement. A portion is given here for guidance regarding its form and content.}
\label{tab:vtss_table}
\begin{tabular}{clrrccccc}
\hline
PN G & Name & RAJ2000 & DEJ2000 & logF(H$\alpha$)& $r_{\rm aper}$    & ~${\rm N}_f$~ & ~c$_{\beta}$~ & Notes \\
\hline
118.0$-$08.6&Vy 1-1&00:18:42.2&53:52:20&$-$10.95$\pm$0.09&4.8&1&~0.38~~&6,10\\
119.3+00.3&BV 5-1&00:20:00.5&62:59:03&$-$11.62$\pm$0.18&5.0&1&$\ldots$&4,5\\  
119.6$-$06.1&Hu 1-1&00:28:15.6&55:57:55&$-$11.00$\pm$0.08&3.2&1&0.44&\\
121.6+00.0&BV 5-2&00:40:21.6&62:51:34&$-$11.60$\pm$0.14&3.2&1&$\ldots$&1,4\\
122.1$-$04.9&Abell 2&00:45:34.7&57:57:35&$-$11.61$\pm$0.10&3.2&2&$\ldots$&6\\
124.3$-$07.7&WeSb 1&01:00:53.3&55:03:48&$-$12.1$\pm$0.2&9.3&1&$\ldots$&\\
126.3+02.9&K 3-90&01:24:58.6&65:38:36&$-$11.90$\pm$0.12&3.2&1&$\ldots$&5,6\\
130.3$-$11.7&M 1-1&01:37:19.4&50:28:12&$-$11.33$\pm$0.10&3.2&1&$\ldots$&\\
130.9$-$10.5&NGC 650/1&01:42:20.0&51:34:31&$-$10.19$\pm$0.05&3.2&1&0.10&\\
138.8+02.8&IC 289&03:10:19.3&61:19:01&$-$10.82$\pm$0.07&3.2&1&1.29&\\
\hline
\end{tabular}
\end{center}
}
\begin{flushleft}
Notes:  (1) Possible PN; (2) pre-PN; (3) transition object; (4) uncertain counts; (5) confused with nearby object; (6) bad pixels in aperture; (7) object near field edge; (8) flux excludes halo; (9) flux corrected for CSPN;  (10) Wolf-Rayet CSPN; (N) previously unpublished object; (V) very low excitation PN; (C) specific comment given.
\end{flushleft}
\end{table*}

However, from a detailed comparison between our VTSS \ha\ data and independent fluxes for 21 bright PNe, we noticed that with this correction, the VTSS integrated fluxes appear to be slightly overestimated (cf. GMR01; F03; Frew 2008).   By fitting a simple linear function to the available data in common, we estimate a new VTSS correction factor of 0.07 dex (in the sense that the VTSS data need to be brightened by 17 per cent).  We applied this correction to all of our VTSS fluxes, except for the objects in fields Aql04 and Ori11 where the original corrections of GMR01 are applied.


Table~\ref{tab:vtss_table} contains the H$\alpha$ fluxes for $\sim$162 PNe measured from the VTSS, brightened by 0.07 dex.  Columns 1 and 2 give the $PN G$ designation and the common name respectively, columns 3 and 4 give the J2000.0 coordinates, while the logarithm of the  \ha\ flux is given in column 5.  The adopted aperture radius in arcmin is given in column 6, the number of measurements from separate fields is given in column 7, the derived extinction in column 7, and any notes are indexed in column 9.

\subsubsection{Comparison of VTSS and literature \ha\ fluxes}\label{sec:flux_comparison_VTSS}
   
We made a comparison of the VTSS \ha\ fluxes with literature data, evaluated in the same way as for SHASSA.  Table~\ref{table:ha_comparison_vtss} presents a statistical comparison between literature \ha\ fluxes and our \ha\ fluxes measured from VTSS after brightening by 0.07 dex.  The columns give the comparison dataset,  the mean $\Delta$log F(\ha) between measurements in the sense VTSS $-$ literature fluxes, the standard deviation of the difference, the number of PNe in common, and the technique and detector used by each data-set.  These can be compared with the comparisons presented in Table~\ref{table:ha_comparison} for SHASSA fluxes.

\begin{table}
\begin{center}
\caption{Statistical comparison of our VTSS \ha\ fluxes with independent literature values. Only data-sets with more than three PNe in common are listed, and the offsets are in the sense VTSS fluxes minus literature. The reference codes are given in the footnotes to Table~\ref{table:ha_comparison}. }
\label{table:ha_comparison_vtss}
\begin{tabular}{lrccclc}
\hline
Reference		&   $\Delta$$F$   & $\sigma$ 				     &     $n$     	&   Meth.      		&     Det.  	 	&     Fig.					\\
\hline
SHASSA			&   	  0.01		& 		0.09			            &	47	        &	Aper 		&      CCD      	&  	\ref{fig:comp3}a	       \\	
\hline
BAR78			&   	   0.03	  	& 		0.06			     	     &	9	       &	Spec  		&    IDS	       		&  	\ref{fig:comp3}b	       \\    
KM81    			&       0.05          	& 		0.03		    	     	     &	4	       &	Aper   		&   PEP	      		  &  \ref{fig:comp3}b		\\  
PTP71			&        0.01	  	& 		0.03			            &	4	       &	Spec  		 &   PEP 			&  	\ref{fig:comp3}b	      \\  
\hline
B01, B03			&   	 $-$0.17	  	& 		0.29				     &	 6     	&	Sp/Ap		 &      CCD            &  	$\ldots$	       \\  %
H97		      		&   	 $-$0.13 		& 		0.14			            &	5	       &	Aper 		 &      CCD            &  	$\ldots$	       \\  %
HGM93		      	&   	 $-$0.03 		& 	       0.23		            	     &	7	       &	Aper 		 &      PCC            &  $\ldots$	       \\  %
K83a   			&    	 0.10		& 	 	0.12		   		     &	9	 	&	Aper		   	&    PEP	      	 	&  	$\ldots$		\\  %
K83b			&      $-$0.03		& 		0.17		   	            &	17	       	&	Aper		 	 &    PEP	     	 	&    $\ldots$		\\  %
KSK90 			&   	0.13	 		& 		0.33	           		     &	14	        &	Spec		&    IDS       		 &  	$\ldots$	        \\  %
RCM05			&   	 0.11	 	& 		0.14			      		&   	4		 &	F-P	 		& CCD	   		 &  	$\ldots$ 		  \\  
\hline
\end{tabular}
\end{center}
\end{table}

\begin{figure*}
\begin{center}
\includegraphics[clip=true,scale=0.54]{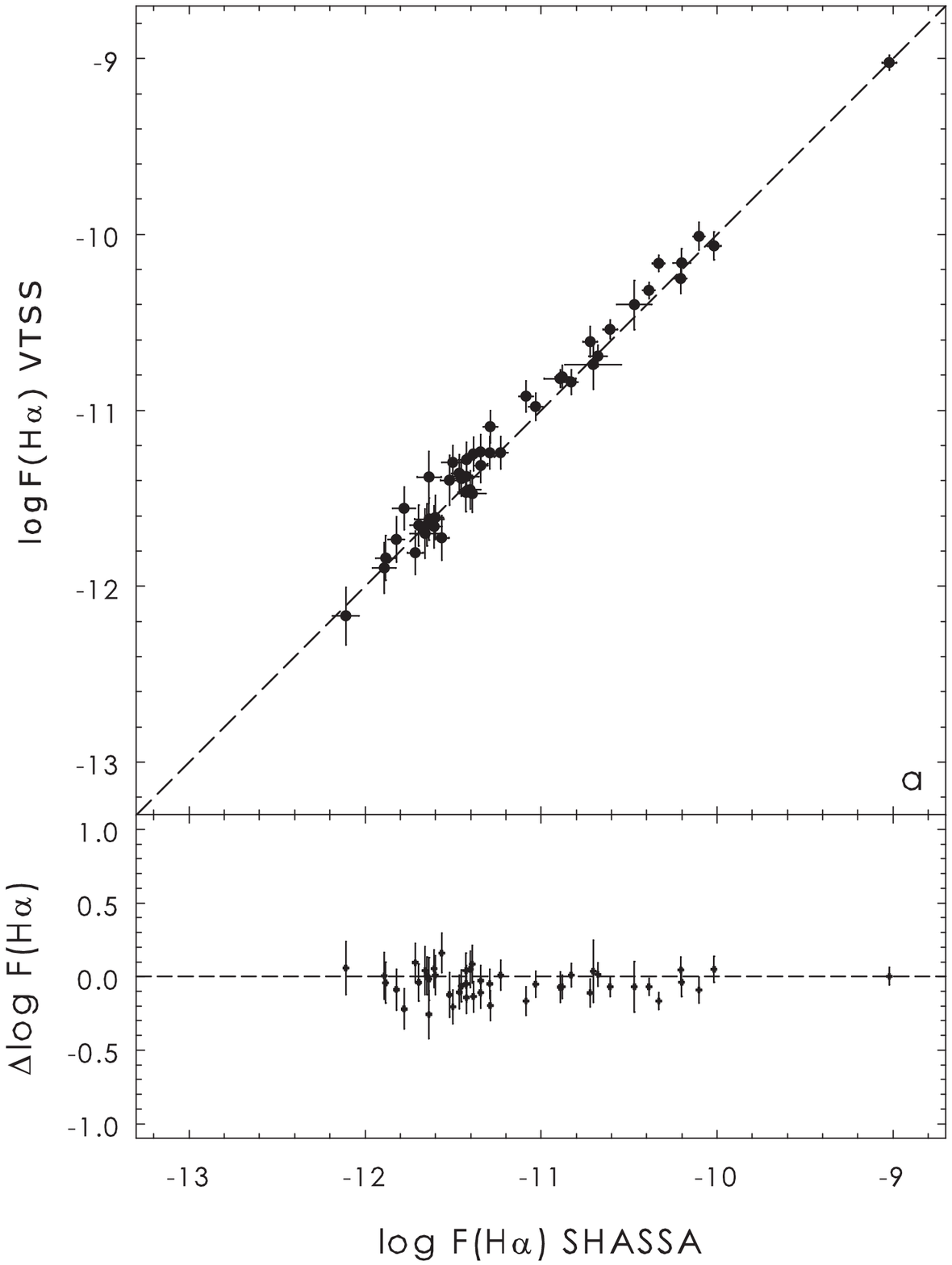}
\includegraphics[clip=true,scale=0.54]{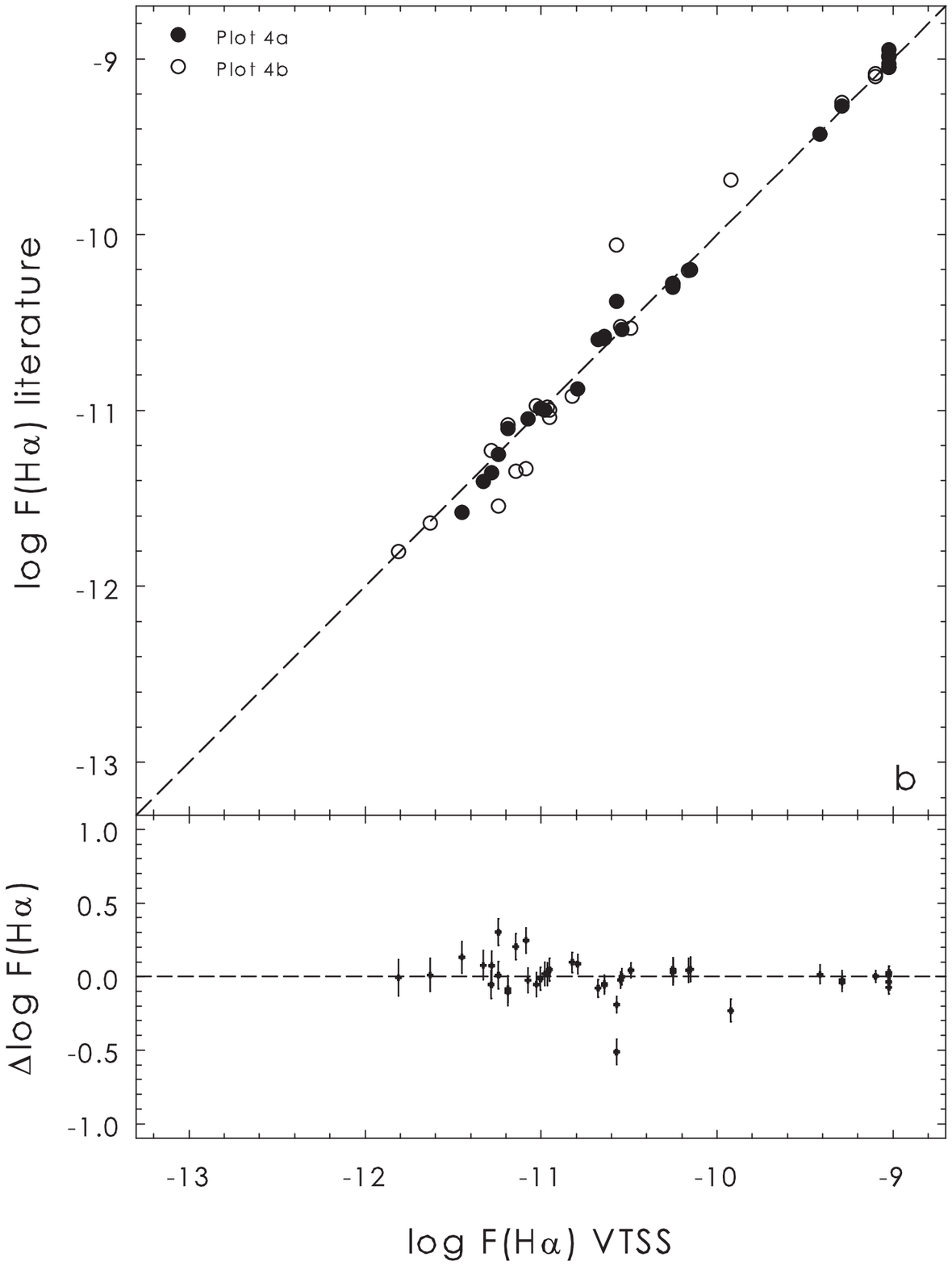}
\caption{The left panel (a) compares our fluxes measured from SHASSA with the corrected VTSS fluxes for 47 PNe in common. The dispersion is very small for fluxes brighter than log\,$F$(\ha) =  $-11.0$.  The right panel (b) compares our VTSS \ha\ fluxes with the most accurate subset of literature fluxes, and the symbol codes refer to the data-sets plotted on Figures~4a and 4b.}
\label{fig:comp3}
\end{center}
\end{figure*}

Figure~\ref{fig:comp3} shows a comparison between our independent VTSS \ha\ fluxes and SHASSA \ha\ fluxes for 47 PNe in common (Figure~\ref{fig:comp3}a;  left panel).    The relation between our VTSS and SHASSA fluxes is linear, with a low dispersion, especially for PNe with log$F$(\ha) $>$$-11.0$.  This independently demonstrates the reliability of our SHASSA fluxes.  In Figure~\ref{fig:comp3}b (right panel) we compare our new VTSS measurements to 179 measurements from 17 independent literature sources which mostly provide small numbers ($<$15) of comparison fluxes.  The abbreviated symbol codes in the right panel are the same as before.  The flux differences vary according to the quality of the published source fluxes, but are generally larger than the SHASSA comparisons plotted in Figure~\ref{fig:comp1}.
We measure a dispersion around the mean between our SHASSA and VTSS fluxes of $\pm$0.09 dex.  Assuming a dispersion in our SHASSA fluxes of 0.04 dex, then the mean VTSS uncertainty can be estimated to be approximately $\pm$0.08 dex.


\section{Extinction determinations}\label{sec:reddenings}

We use our homogenous catalogue of \ha\ fluxes to calculate independent extinction determinations for a subset of $\sim$270 PNe with reliable integrated \hb\ fluxes in the literature.  We take these from Copetti (1990), ASTR91, CKS92, DH97 and WCP05, and exclude any with an uncertainty in log$F$(\hb) of $>$0.06 dex.  We also exclude any objects with an uncertainty in \ha\ of $>$0.08 dex.   The logarithmic extinction at H$\beta$, $c_{\beta}$ = ${\rm log}F(H\beta) - {\rm log}\,I(H\beta$), is derived for each PN by comparing the observed Balmer decrement with the theoretical one for case B recombination.  In this study, we have used the $R = 3.1$ Galactic reddening law of Howarth (1983) and adopt an intrinsic line ratio of $I$(H$\alpha$)/$I$(H$\beta$) = 2.86 (Hummer \& Storey 1987), or log\,$I$(\ha/\hb) = 0.456.  This value is appropriate for T$_e$ = 10$^4$\,K and the range of densities seen in most PNe.  The logarithmic extinction is then given by the following expression:



\begin{equation}
\label{reddening}
c_{\beta} =  3.125 \times {\rm log}\left[{F({\rm H}\alpha) \over F({\rm H}\beta)}\right] - 1.43
\end{equation} 

where as before, $F$(\ha) and $F$(\hb) are the observed integrated fluxes.  The logarithmic extinction at \ha, $c_{\alpha}$, can also be derived using:

\begin{equation}
\label{reddening_alpha}
c_{\alpha} = 0.70 \times c_{\beta}
\end{equation} 

Our \hb\ logarithmic extinctions should be accurate to better than 0.10~dex, and should not be affected by the unknown intensity of the \HeII\ Pickering lines which contribute to the \ha\ and \hb\ fluxes, as to first order, the intrinsic Pi\,6/Pi\,8 ratio of $\sim$2.7 is nearly identical to the intrinsic \ha/\hb\ ratio (Hummer \& Storey 1987).    The calculated extinctions are given in column~11 of Table~\ref{tab:shassatable} and column~9 of Table~\ref{tab:vtss_table}.   There were a few formally negative values for the extinction arising from measurement errors in both the  \ha\ and \hb\ fluxes, and these have been reset to zero in the tables.   

Our extinction determinations are generally in very good agreement with other values taken from the literature (Pottasch et al. 1977; Feibelman 1982; Kaler \& Lutz 1985; Gathier, Pottasch \& Pel 1986; CKS92; Stasi\'nska et al. 1992; Tylenda et al. 1992; Giammanco et al. 2011) and with the total line-of-sight extinction to each object (Schlafly \& Finkbeiner 2011).  In a follow-up paper, we will undertake a comprehensive analysis of the optical and radio extinctions for a large number of Galactic PNe (Boji\v{c}i\'c et al. 2011a; Boji\v{c}i\'c et al., in preparation).

\section{Future Work}\label{sec:future}

This paper is the first in an ambitious project which plans to determine an integrated hydrogen-line flux for almost all of the currently-known Galactic PNe, and the new discoveries coming on-stream (e.g. Parker \& Frew 2011), especially from the IPHAS survey.  It is important to note that the SHASSA survey only covers the southern hemisphere, and that the VTSS is unlikely to be reborn. Furthermore,  these surveys are of low angular resolution and cannot provide reliable fluxes nor, in many cases, detections for faint PNe in complex regions of high background emission.  Alternative sources for determining \ha\ fluxes for these PNe are required.  

In this vein, Parker et al. (2012c) demonstrated that the SHS survey, though photographic in origin, can never the less be used to provide flux estimates for resolved nebulae.  As part of a pilot study, moderate-precision integrated \ha\ fluxes for $\sim$50 faint PNe were measured.  This is important as the 900 or so MASH PNe not detected in SHASSA were all discovered on the SHS.   We are now extending this pilot to undertake a larger project to determine \ha\ fluxes for  a large number of MASH PNe that have no short-term prospects of alternative flux measurements.   These include the faintest PNe detectable by the SHS including those with an interstellar $V$-band extinction exceeding 10 magnitudes (see Parker et al. 2012b).  The SHS should extend the flux limit for Galactic PNe down to log$F$(\ha) $\simeq-15$, or a factor of a few hundred fainter than the SHASSA limit.  

In the near future, many more global \ha\ fluxes will be determined for PNe from the IPHAS survey (Drew et al. 2005), within 5\arcdeg\ of the northern Galactic plane.  The digital IPHAS images, like the SHS data, are not only a powerful discovery medium, but have the advantage of providing accurate \ha\ fluxes for the hundreds of PNe located within the 1800 square degree footprint.  The initial IPHAS point source catalogues (Gonz\'ales-Solares et al. 2008; Witham et al. 2008) hold few compact PNe, but many more extended nebulae are being discovered from visual examination of the imaging data (e.g. Viironen et al. 2009b, 2011; Sabin et al. 2010).  A uniform flux calibration across the whole survey, accurate to $\simeq$0.01\,dex, is in progress.  This will allow the determination of precise \ha\ fluxes for many northern extended PNe.  To date, only a handful of \ha\ fluxes for PNe from the IPHAS survey have been published (Mampaso et al. 2006; Wesson et al. 2008; Viironen et al. 2009a, 2011; Corradi et al. 2011).   

The southern counterpart of the digital IPHAS Survey is the new VST/OmegaCam Photometric \ha\ Survey of the Southern Galactic Plane (VPHAS+; Drew \& Raddi 2012; J. Drew et al., in preparation)\footnote{http://www.vphasplus.org/}.  This survey commenced in 2012 using the recently completed VLT Survey Telescope, equipped with OmegaCam (Kuijken et al. 2004).  VPHAS+ will have a depth about two magnitudes fainter than the SHS for point sources, and comparable sensitivity to diffuse emission.  Besides finding new PNe missed by the SHS, VPHAS+ should provide accurate \ha\ fluxes for many southern PNe below the flux limit of SHASSA.  However, its areal coverage is only $\sim$50 per cent of the SHS survey, so the SHS remains the only \ha\ survey able to cover most of the southern Galactic plane at good sensitivity and resolution for the foreseeable future.

We also note the Palomar Transient Factory (PTF; Rau et al. 2009), a series of wide-field surveys that will investigate the variable sky on time scales from minutes to years.  Amongst many science goals, a set of narrowband (\ha, \ha-off and \OIII) surveys will be conducted, covering the sky north of $\delta$ = $-28$\arcdeg. The estimated 5$\sigma$ intensity limit for the PTF \ha\ survey will be $\sim$0.6\,R, allowing the discovery of new faint PNe, and providing new integrated fluxes for faint PNe and other emission-line objects.

We also plan to derive a set of independent \ha\ fluxes from narrowband archival images taken with HST and the F656N filter.  Few of these fluxes have been published to date (e.g. Sahai, Nyman \& Wootten 2000; Wesson \& Liu 2004; Miszalski et al. 2011), so any further data will be worthwhile.   

Our catalogue is complementary to fluxes obtained from imaging surveys at other wavelengths, such as the very bright \OIII\ line at $\lambda$5007\AA\  (e.g. Kovacevic et al. 2011), or in the light of \SIII\ $\lambda$9532\AA\ (Kistiakowsky \& Helfand 1993, 1995; Jacoby \& Van de Steene 2004; Shiode et al. 2006).  With the exception of the VLE PNe, the \SIII\ $\lambda$9532\AA\ line essentially becomes the brightest apparent line when the total $V$-band extinction is between 5 and 12 mag (Van de Steene, Sahu \& Pottasch 1996; Jacoby \& Van de Steene 2004).  If NIR spectra are available, then the integrated flux in an atomic hydrogen line such as Paschen-$\gamma$ can then be bootstrapped from the observed $\lambda$9532\AA\ flux.  

For the many PNe expected to be completely hidden in the optical due to very high extinction, global fluxes in a longer wavelength hydrogen line will be needed.  Brackett-$\gamma$ is recommended as it one of the brightest lines accessible from the ground when the visual extinction exceeds $\sim$12 magnitudes (Jacoby \& Van de Steene 2004; Jacoby et al. 2010); at this level of extinction H$\alpha$ is rapidly becoming undetectable.\footnote{Except for the most compact PNe, the \hb\ line becomes undetectable when the V-band extinction exceeds 8--9 mags.}  Brackett-$\alpha$  (e.g. Forrest et al. 1987) might also be a useful secondary choice for very reddened PNe.  The intrinsically brighter Paschen-$\alpha$ line is only accessible from space, but will prove useful when available (e.g. Wang et al. 2010; Dong et al. 2011).

\section{Summary}\label{sec:summary}

We present reliable \ha\ fluxes for 1258 unique PNe, including fluxes for 1234 PNe measured from SHASSA and VTSS images.  This is the largest compilation of homogeneously derived fluxes yet measured. In Appendix~A, we list the corrected \ha\ fluxes for 49 PNe taken from the literature, including 24 PNe not detected on our SHASSA or VTSS images, all set to a common zero-point.  

Aperture photometry on the digital images was performed to extract integrated \ha+\NII\ fluxes in the case of SHASSA, and \ha\ fluxes from VTSS. The  \NII\ contribution was then deconvolved from the SHASSA flux using spectrophotometric data taken from the literature or derived by us.  Comparison with previous work shows that our flux scale has no significant zero-point error.   Including literature flux measurements for  another 200 PNe that were either too faint, or fell outside of the combined survey coverage, there are now reliable integrated \ha\ fluxes for nearly 45\% of the Galactic PN population.  We also list new independent reddening determinations for $\sim$270~PNe, derived from a comparison of our \ha\ data with the best literature \hb\ fluxes, and in Appendix~B, we provide integrated  \ha\ fluxes for another 76 nebulae formerly classified as PNe.  

Along with providing the community with an legacy resource with many applications, our benchmark catalogue of \ha\ fluxes can be used to calculate surface-brightness distances for each PN (Frew 2008), estimate new Zanstra temperatures for PN central stars with accurate photometry (e.g. De Marco et al. 2012), provide baseline data for photoionization modelling, and allow a comparison with data at radio (Condon \& Kaplan 1998; Boji\v{c}i\'c et al. 2011a), mid-infrared (e.g. Chu et al. 2009; Zhang, Hsia \& Kwok 2012), near-IR (Guerrero et al. 2000; Schmeja \& Kimeswenger 2001; Speck et al. 2002; Hora et al. 2006;  Cohen et al. 2011; Froebrich et al. 2011) and ultraviolet and X-ray wavelengths (Bianchi 2012; Kastner et al. 2012) in order to better understand the physical processes occurring within PNe.   Lastly, we will determine new absolute \ha\ magnitudes for delineating the faint end of the PNLF (Ciardullo 2010).  Such an \ha\ PNLF can be compared with the widely-used  \OIII\ $\lambda$5007\AA\  PNLF, both within and outside the Milky Way (e.g. M\'endez et al. 1993; Jacoby \& De Marco 2002; Sch\"onberner et al. 2007; Reid \& Parker 2010; Kovacevic et al. 2011; Ciardullo 2012).    


Ultimately, we hope to conduct a multi-wavelength statistical analysis of the Galactic PN population in a way that has not yet been achieved.   This analysis will be crucial to our overall understanding of the origin of PNe, including the role of binary companions in their shaping and evolution (De Marco 2009; De Marco \& Soker 2011), the physics of AGB mass loss, and the return of metals to the ISM.  This paper is a first step towards this goal.

\section*{ACKNOWLEDGEMENTS}
We thank the referee, C.R. O'Dell, for comments which helped to improve the paper.  D.J.F. thanks Macquarie University for a MQ Research Fellowship and I.S.B. is the recipient of an Australian Research Council Super Science Fellowship (project ID FS100100019).  This research has made use of the SIMBAD database and the VizieR service, operated at CDS, Strasbourg, France, and also utilized imaging data from the Southern \ha\ Sky Survey Atlas (SHASSA) and the Virginia-Tech Spectral Line Survey (VTSS), which were produced with support from the National Science Foundation (USA).  Additional data were used from the AAO/UKST \ha\ Survey, produced with the support of the Anglo-Australian Telescope Board and the Particle Physics and Astronomy Research Council (UK), and the Wisconsin H-Alpha Mapper (WHAM), produced with support from the National Science Foundation.  This project also made use of APLpy, an open-source plotting package for Python.   We thank Greg Madsen for his help with the WHAM data, and Warren Reid for his assistance in measuring the line fluxes from our unpublished spectra.

\appendix

\section{Recalibrated \ha\ fluxes}\label{appendix_recal}

Other potentially useful integrated \ha+\NII\ fluxes have been published in the literature, especially for several large, evolved PNe.   In particular,  Xilouris et al. (1996; XPPT) presented \ha+\NII\ and \OIII\ fluxes for eight evolved PNe (see also Papamastorakis, Xilouris  \& Paleologou 1994), while Xilouris et al. (1994) and Ali et al. (1997) presented integrated \ha+\NII\ fluxes for another five PNe. 
However, we found that the XPPT fluxes to be inconsistent with other measurements. A comparison between their quoted surface brightness data and independent SHASSA measurements show the XPPT surface brightness data to be accurate, but the integrated fluxes were found to be too faint by a factor of $\sim$20, presumably the result of a simple reduction error.  In order to make a comparison with other sources, firstly H$\alpha$ fluxes were derived from their quoted `red' fluxes by deconvolving the \NII\ emission for each PN, using the references from Section~\ref{sec:spectrophotometric}.  The characteristics of the broad \ha+\NII\ filter ($\lambda_{\rm eff}$ = 6570\AA, FWHM = 75\AA) are described by XPPT and Mavromatakis et al. (2000) which indicates the transmission for the \NII\ doublet is identical to H$\alpha$ making the deconvolution calculation straightforward (i.e. $K_{\rm tr}$ = 1; see Table~\ref{table:filter_constants}).  We then compared the \ha\ fluxes from our SHASSA and other literature data with the XPPT fluxes; they were found to be offset in the mean by 1.32 $\pm$ 0.08 dex ($n$ = 7) and this correction was hence applied to all their original fluxes.  

Table~\ref{corrected_fluxes} gives the original \ha+\NII\ fluxes adopted from these sources and our derived \ha\ fluxes.  
The {[\rm N \sc ii]}/H$\alpha$ ratios were taken from the references described in Section~\ref{sec:spectrophotometric} and we used the filter transmission coefficients from Table~\ref{table:filter_constants}, and the principles adopted previously.  The \ha\ fluxes from XPPT corrected for the zero-point offset are presented in the last column.

Another set of useful emission-line data in \ha, \NII\ and \OIII\ is presented by Gieseking, Hippelein \& Weinberger (1986; GHW) and Hippelein \& Weinberger (1990; HW90).  These authors used a Fabry-Perot spectrometer using 1\arcmin\ or 2\arcmin\ apertures (i.e. smaller than most of the PNe they studied), presenting surface brightness values (given in mag per arcsec$^{2}$) for each PN.   These surface brightness values were not immediately reconcilable with the fluxes of XPPT and Kaler (1983b) for the evolved PNe in common, nor with our SHASSA data, as the adopted formula to convert from magnitudes to cgs units is not given by either GHW nor HW90.  However, a comparison of the surface brightness measurements of GHW for IsWe~1 and IsWe~2 with the absolute surface fluxes for these same PNe presented by Ishida \& Weinberger (1987), allowed the determination of the transformation equation used, which is that given by Pottasch (1984), viz:

\begin{equation}
\label{eq:mag-conv}
m_{\lambda} = -2.5 \,{\rm log} F(\lambda) - 15.77
\end{equation}

where $\lambda$ is the emission line species of interest. Consequently the same zero-point was used for \OIII, \NII\ and \ha, despite the range in wavelength of these emission lines (cf. Allen 1973, p. 197).  Using equation~\ref{eq:mag-conv}, the surface brightnesses in \ergcms arcsec$^{-2}$ in each  line were then determined.  However, to get the integrated flux, one needs to know the angular size of these PNe.  The dimensions were generally taken from Frew (2008).   Table~\ref{Hippelein_corr} summarises the adopted dimensions and derived \ha\ fluxes for these PNe.

\begin{table}
{\footnotesize
\begin{center}
\caption{Summary details of the \ha+\NII\ filters used in various studies, and the adopted transmission coefficients used in Equation~\ref{eq:shassa_deconvolve}. These were calculated from the filter transmission characteristics given in the references.}
\label{table:filter_constants}
\begin{tabular}{lccc}\hline
Reference    	 		&$\lambda_{\rm eff}$ (\AA) 	&  FWHM  (\AA)   	 	&	$K_{\rm tr}$	\\
\hline
SHASSA				&       6563     			&   32     		            		&	0.375		\\
VTSS				&       6566     			&   17.5     		            		&	0.00			\\
SHS					&       6590    			&   70     		            		&	0.725		\\
\hline
APS97				&       6574     			&   104     	            			&	1.00			\\
JaSt					&       6569     			&   80      	          	 		&	1.00			\\
XPS94            			&       6570     			&   75      	          	 		&	1.00			\\  
XPPT	          		&       6570     			&   75      	          	 		&	1.00			\\
\hline
\end{tabular}
\end{center}
}
\end{table}

\begin{table*}
\begin{center}
\caption{Integrated \ha+\NII\ fluxes taken from Xilouris et al. (1994), Xilouris et al. (1996; XPPT) and Ali et al. (1997).  The XPPT fluxes have been further corrected as described in the text. The flux uncertainties are estimated to be $\pm$\,0.08 dex.}\label{corrected_fluxes}
\begin{tabular}{llcccclc}
\hline
Name~~~~~~~~~& 	PN G		& $R_{\rm \NII}$  	& ~log$F_{\rm red}$~    & log$F$(H$\alpha$)	&log$F$(H$\alpha$)	&  ~Ref. 			&  	Notes		\\
              		&					& 				&    	                 		&                  			&      (corrected)     	 &			 	&  			\\
\hline
Abell 4   		&    144.3$-$15.5~~~	&   0.28	   		&   $ -11.80$	       		&   $-11.91$   			&	$-11.91$       		&    APS97~~		&  	...			\\
Abell 7     	&	215.5$-$30.8  		&   0.9       		&   $-11.59$  	     		&   $-11.85$        		&    $ -10.58 $      		&     XPPT		&  ...				\\  
Abell 8   		&     167.0$-$00.9		&   0.97	    		&   $ -11.60$  			&   $-11.90$ 			&     $-11.90$ 	 		&   	APS97		&  ...				 \\ 
Abell 62    	&	047.1$-$04.2 		&   1.5     		&   $-11.96$ 			&   $ -12.29$     		&    $ -11.02$       		&     XPPT		&  ...				 \\ 
Abell 74    	&	072.7$-$17.1		&  1.3    			&   $ -11.51$ 	 		&$ -11.80$        		&    $ -10.52$   		&     XPPT		&  ...			 	\\ 
G4.4+6.4   	&     004.3+06.4 		&   2.4    			&   $ -10.70$       		&  $-11.23$ 			&  	$-11.23$        		&     XPS94		& 	a			\\
HFG 1      	&	136.3+05.5 		&  0.4      		&   $ -11.59$  			&   $-11.77 $     		&    $ -10.49$   		&     XPPT	 	&  	...			 \\
IsWe 1     	&	149.7$-$03.3 		&   0.75    		&   $ -11.96$  			& $-12.07$       		&    $ -10.80$        		&     XPPT		&  	...			 \\
IsWe 2     	&	107.7+07.8 		&   1.45    		&   $ -11.37$  			& $ -11.76$           		&    $ -10.48$     		&     XPPT		&  	...			 \\
J 320   		&     190.3$-$17.7 		&   0.04	    		&     $ -10.80$  		&   $-10.82$ 			&    	$-10.82$      		& 	APS97		&  	...			\\
\noalign{\smallskip}
M 2-55  		&     116.2+08.5 		&   1.7	    		&     $ -10.74$  		&   $-11.17$ 			&	$-11.17$	  		&     APS97 		&  ...				\\
Sh 2-68    	&	030.6+06.2 		&   0.5    			&   $ -11.66$  	 		& $ -11.84$       		&    $  -10.56$   		&     XPPT	  	&  	a,b			\\
Sh 2-176   	&	120.2$-$05.3		&   2.3      		&   $-11.41$  			& $-11.93$        		&    $-10.65 $       		&     XPPT	   	&  	...			\\
Sh 2-188   	&	128.0$-$04.1		&   1.9      		&   $ -10.87$   		&  $-11.35$       		&    $ -10.07 $     		&     XPPT		&  ...				\\
\hline
\end{tabular}
\end{center}
{\footnotesize
{\flushleft Notes:~~  (a)  PN status uncertain; (b) flux excludes outer halo.\\}
}
\end{table*}

\begin{table*}
\begin{center}
\caption{Integrated \ha\ fluxes derived from the surface brightness data presented by Gieseking, Hippelein \& Weinberger (1986, GHW), Hippelein \& Weinberger (1990, HW90), and Saurer \& Weinberger (1987, SW87), who utilised the MPI Fabry-Perot Interferometer (Hippelein \& M\"unch 1981).  For completeness, we also quote the integrated \ha\ fluxes taken with this instrument for two smaller PNe (Saurer 1997a,b).  The flux uncertainties are estimated to be $\pm$\,0.05 dex for the smaller objects up to $\pm$\,0.25 dex for the largest.}
\label{Hippelein_corr}
\begin{tabular}{llcccclc}
\hline
Name~~~~~~~~&	PN G 			&~Size~~		&	$S$(\ha)				&	log$S$(\ha)	&	log$F$(\ha) 					&	~Ref.		&  Notes			 \\
			 &					&	(\arcsec)	& (mag/$\square$\arcsec)	& (erg/cm$^2$/s/$\square$\arcsec) & (erg/cm$^2$/s)		&                   	&				\\
\hline																									
Abell 28		&	158.8+37.1~~~	&	320			&	25.1	 &		$-12.79$	&	$-11.44$				&	HW90~~		&  ...			\\
Abell 34		&	248.7+29.5		&	292			&	24.1	 &		$-12.39$	&	$-11.12$				&	HW90		&  ...			\\
Abell 39		&	047.0+42.4 		&	174			&	23.9	 &		$-12.31$	&	$-11.49$				&	HW90		&  ...			\\
Abell 61		&	077.6+14.7 		&	199			&	23.6	 &		$-12.19$	&	$-11.26$				&	GHW		& ... 			\\
Abell 62		&	047.1$-$04.2		&	160			&	22.8	 &		$-11.87$	&	$-11.12$				&	HW90		&  ...			\\
Abell 71		&	084.9+04.4 		&	157			&	22.0	 &		$-11.55$	&	$-10.82$				&	HW90		&  ...			\\
Abell 74		&	072.7$-$17.1		&	795			&	24.5	 &		$-12.55$	&	$-10.41$				&	GHW		& ... 			\\
DHW 5		&	111.0+11.6		&	595			&	24.3	 &		$-12.47$	&	$-10.58$				&	GHW		&  a				\\
EGB 6		&	221.5+46.3 		&	780			&	26.3	 &		$-13.27$	&	$-11.15$				&	HW90		&  ...			\\
IsWe 1		&	149.7$-$03.3		&	745			&	25.6	 &		$-12.99$	&	$-10.91$				&	GHW		&  ...			\\
\noalign{\smallskip}
IsWe 2		&	107.7+07.8 		&	970			&	25.1	 &		$-12.79$	&	$-10.48$				&	GHW		&  ...			\\
JnEr 1		&	164.8+31.1 		&	380			&	23.5	 &		$-12.15$	&	$-10.65$				&	HW90		&  ...			\\
K 3-82		&	 093.3-00.9		&	20			&	...	 &			...	&	$-11.60$				&	S97b		&  ...			\\
M 1-79		&	093.3-02.4 		&	33			&	...	 &			...	&	$-11.04$				&	S97a		&  ...			\\ 
PuWe 1		&	158.9+17.8 		&	1210			&	25.2	 &		$-12.83$	&	$-10.33$				&	GHW		&  ...			\\
SaWe 3		&	013.8$-$02.8 		&	95			&	...	 &		$-11.58$	&	$-11.19$				&	SW87		&  b				\\  
Sh 2-68		&	030.6+06.2		&	410			&	24.0	 &		$-12.35$	&	$-10.79$				&	HW90		&  c,d			\\
We 1-10		&	086.1+05.4		&	190			&	24.0	 &		$-12.35$	&	$-11.46$				&	HW90		& ... 			\\
We 3-1		&	044.3+10.4 		&	166			&	23.2	 &		$-12.03$	&	$-11.25$				&	HW90		&  ...			\\
\hline
\end{tabular}
\end{center}
{\footnotesize 
{\flushleft Notes:~~  (a)  HII region;  (b) the $S$(\ha) and total $F$(\ha) values are derived from the measured \ha\ flux of $5.5 \times 10^{-13}$ \ergcms\ observed through a 31\arcsec\ aperture;  (c) PN status uncertain; (d)  Flux excludes outer halo.\\}
}
\end{table*}

There are a few other moderately-sized PNe from Abell (1966) which have no flux data at present, either because they are too faint for SHASSA and VTSS, or because they are located outside the bounds of the available survey fields. Since any flux data on these poorly-studied objects are welcome, a re-investigation of the Abell (1966) photographic data is warranted. Nebular magnitudes were estimated from the POSS~I survey blue and red plates.    The nebular surface brightness of a PN was compared with spots of different densities on film strips exposed with a sensitometer, in turn calibrated with extra-focal images of standard stars.  Further details on the method are given in Abell (1966).  

From the calibrated surface brightness and the angular size of the nebula in each passband, integrated photographic and photo-red magnitudes were determined for each object, but we are only concerned with the photo-red magnitudes here.
These were adjusted following Jacoby (1980) and then corrected here for the contribution of the nitrogen lines, assuming equal throughput for the \ha\ and \NII\  lines from the broadband Plexiglass filter used in conjunction with red-sensitive (Kodak 103a-E) photographic plates. The conversion to `\ha' magnitudes is given by:

\begin{equation}
\label{eq:abell_mags}
m_{({\rm H}\alpha)} = m_{\rm pr}  - 2.5\,{\rm log} \left( \frac{1}{R_{\rm [N\,II]} + 1} \right)
\end{equation}

where $R_{\rm [N\,II]}$ is defined as before.  The resulting `\ha' magnitudes were then compared with the mean of all available integrated \ha\ fluxes from our database. A least-squares linear fit to the flux-magnitude relation gave:

\begin{equation}
\label{eq:abell_fluxes}
m_{({\rm H}\alpha)}  =  -2.5\,{\rm log}\,F({\rm H}\alpha) - 14.35
\end{equation}

which can be compared with the expressions given by Ciardullo (2010) and Reid \& Parker (2012).  Equation~\ref{eq:abell_fluxes} was used to determine approximate \ha\ fluxes for all the PNe with $m_{\rm pr}$ magnitudes  listed in Abell (1966) that have \NII/\ha\ ratios available.  We compared the derived fluxes with literature \ha\ data, and with our SHASSA and VTSS\ha\ fluxes, finding a dispersion of the Abell `\ha' fluxes compared to  our independent SHASSA and VTSS fluxes of $\pm$0.22\,dex ($n$ = 44) and $\pm$0.21\,dex ($n$ = 20) respectively.  The Abell `\ha' fluxes derived in this way are surprisingly good, and better than some modern CCD studies (see Table~\ref{table:ha_comparison}).  Rather then listing the full list of  derived \ha\ fluxes, we present in Table~\ref{tab:abell_ha_fluxes}  data for the 14 PNe and two mimics which either have no measurements from elsewhere in this paper, or have inconsistent fluxes in the literature. 


\begin{table*}
\begin{center}
\caption{Magnitudes and derived H$\alpha$ fluxes for 23 PNe and two mimics from Abell (1966) following the method described in the text. The flux uncertainties are estimated to be $\pm$\,0.22 dex.}
\label{tab:abell_ha_fluxes}
\begin{tabular}{lccccccc}
\hline
Name &PN G   &Other &  $R_{\rm \NII}$ &  m$_{\rm pr}$& m$_{({\rm H} \alpha)}$& F(H$\alpha$) &Note \\
\hline
Abell 1		&	119.4+06.5		&	...	&	0.7	&	15.7	&	16.3	&	$-12.21$	&	a	\\
Abell 3		&	131.5+02.6		&	...	&	0.4	&	14.3	&	14.6	&	$-11.61$	&	...	\\
Abell 5		&	141.7$-$07.8	&	...	&	4.0	&	13.7	&	15.4	&	$-11.98$	&	...	\\
Abell 6		&	136.1+04.9		&	...	&	0.2	&	13.6	&	13.8	&	$-11.26$	&	...	\\
Abell 8		&	167.0$-$00.9	&	...	&	1.0	&	15.0	&	16.0	&	$-12.00$	&	...	\\
Abell 9		&	 172.1+00.8	 	&	...	&	1.6	&	16.3	&	17.3	&	$-12.69$	&	...	\\
Abell 19	&	200.7+08.4		&	...	&	1.7	&	15.8	&	16.9	&	$-12.43$	&	...	\\
Abell 30	&	208.5+33.2		&	...	&	0.1	&	14.5	&	14.6	&	$-11.58$	&	...	\\
Abell 33	&	238.0+34.8		&	...	&	0.2	&	12.1	&	12.3	&	$-10.68$	&	...	\\
Abell 49	&	027.3$-$03.4	&	...	&	1.3	&	14.2	&	15.1	&	$-11.80$	&	...	\\
\noalign{\smallskip}
Abell 54	&	055.3+06.6		&	...	&	1.2	&	15.5	&	16.3	&	$-12.29$	&	...	\\
Abell 55	&	033.0$-$05.3	&	...	&	1.0	&	13.0	&	13.8	&	$-11.26$	&	...	\\
Abell 59	&	053.3+03.0		&	...	&	2.1	&	13.2	&	14.4	&	$-11.52$	&	...	\\
Abell 60	&	025.0$-$11.6	&	...	&	0.0	&	14.7	&	14.7	&	$-11.64$	&	...	\\
Abell 69	&	076.3+01.1		&	...	&	5.2	&	15.3	&	17.3	&	$-12.67$	&	a	\\
Abell 72	&	059.7$-$18.7	&	...	&	0.1	&	13.8	&	13.9	&	$-11.30$	&	...	\\
Abell 73	&	095.2+07.8		&	...	&	1.3	&	14.7	&	15.6	&	$-11.99$	&	...	\\
Abell 75	&	101.8+08.7		&~NGC 7076~&	0.0	&	15.2	&	15.2	&	$-11.85$	&	...	\\
Abell 77	&	097.5+03.1		&Sh 2-128&	0.2	&	12.5	&	12.7	&	$-10.83$	&	b	\\  
Abell 78	&	081.2$-$14.9	&	...	&	0.1	&	15.3	&	15.4	&	$-11.91$	&	...	\\
\noalign{\smallskip}
Abell 79	&	102.9$-$02.3	&	...	&	6.9	&	12.1	&	14.4	&	$-11.49$	&	...	\\
Abell 81	&	117.5+18.9		&	IC 1454	&	0.3	&	13.6	&	13.9	&	$-11.32$	&	...	\\
Abell 83	&	113.6$-$06.9	&	...	&	0.8	&	15.9	&	16.6	&	$-12.31$	&	...	\\
Abell 85	&	...				& CTB 1	&	0.6	&	8.6	&	9.1	&	$-9.5:  $		&	a,c	\\  
Abell 86	&	118.7+08.2 		&	...	&	0.0:	&	14.8	&	14.8	&	$-11.67$	&	a	\\
\hline
\end{tabular}
\end{center}
{\flushleft  Notes: (a) No other integrated flux determination in the literature; (b) \HII\ region; (c) supernova remnant.}
\end{table*}

\section{\ha\ fluxes for misclassified nebulae}\label{mimic_fluxes}

Many different kinds of objects can masquerade as PNe in extant catalogues (see FP10 for a detailed review).  These objects include compact \HII\ regions (e.g. Kimeswenger 1998; Copetti et al. 2007; Cohen et al. 2011; Anderson et al. 2012), Str\"omgren zones around low-mass stars (Reynolds 1987; Hewett et al. 2003; Chu et al. 2004; Madsen et al. 2006; Frew 2008; Frew et al. 2010; De Marco et al. 2012; cf. Wareing 2010), ejecta shells around Wolf-Rayet and other massive stars (e.g. Duerbeck \& Reipurth 1990; Cohen \& Barlow 1975; Crawford \& Barlow 1991a.b; Chu 2003; Egan et al. 2002; Morgan, Parker \& Cohen 2003;  Cohen et al. 2005a; Stock \& Barlow 2010), B[e] and related stars (e.g. Lamers et al. 1998),  symbiotic systems (Blair et al. 1983; Lutz 1984; Acker, Lundstr\"om \& Stenholm 1988; Corradi 1995;  Belczy\'nski et al. 2000; Corradi et al. 2008, 2010; Lutz et al. 2010), Herbig-Haro objects (Cant\'o 1981; Cappellaro et al. 1994), evolved supernova remnants (e.g. Mavromatakis et al. 2001a,b;  Stupar et al. 2007, 2008; Sabin et al. 2012), as well as old nova shells and bow-shock nebulae (see FP10, and references therein).   

For reviews of the problem, the reader is referred to Acker \& Stenholm (1990), Kohoutek (2001), Parker et al. (2006), Cohen et al. (2007, 2011), Kwok (2010), FP10, and Frew \& Parker (2011), while individual lists of misclassified PNe have been published by Sabbadin (1986), Acker et al. (1987), Acker \& Stenholm (1990), Zijlstra, Pottasch \& Bignell (1990), Kohoutek (2001), Frew (2008), and Miszalski et al. (2009).  Since our input database drew on the older PN catalogues (Acker et al. 1992, 1996; Kohoutek 2001), we measured  \ha\ fluxes for 76 objects of various kinds that were formerly classified as  PNe.  

We do not provide an exhaustive list of mimics but only include those 76 objects that were cleanly detected with our pipeline.  Nonetheless, many well-known objects are included, such as the symbiotic outflow Hen~2-104 (Lutz et al. 1989; Corradi \& Schwarz 1993; Santander-Garc\'ia et al. 2008), the WR nebula M~1-67 (Cohen \& Barlow 1975; Crawford \& Barlow 1991a; Esteban et al. 1991; Grosdidier et al. 2001), and Kepler's supernova remnant (Leibowitz \& Danziger 1983), for which integrated \ha\ fluxes should be of interest to the wider community.  
Table~\ref{tab:mimic_table} lists the \ha\ fluxes for these objects.  Columns 1 and 2 give the $PN G$ designation and the common name respectively, columns 3 and 4 give the J2000.0 coordinates, while the adopted value of $R_{\NII}$, the logarithm of the red flux and the logarithm of the \ha\ flux are given in column 5, 6 and 7.  The adopted aperture radius in arcmin is given in column 8, the number of measurements from separate fields is given in column 9, the classification of the object in column 10, and any notes are indexed in column 11.    Objects that are likely to be mimics, but which are still considered to be PNe by some authors are kept in the main tables for now, and are flagged accordingly (see Tables~\ref{tab:shassatable} and \ref{tab:vtss_table}).  

The abbreviations used for the classification are as follows: \HII: \HII\ region; ELS: emission-line star; SNR: supernova remnant; BCD: blue compact dwarf; SyS: symbiotic star; post-RSG: post-red supergiant; WR neb:  Wolf-Rayet nebula; (c)LBV: (candidate) luminous blue variable; CV: cataclysmic variable.

\clearpage

\onecolumn
\newpage

\begin{center}
{\footnotesize
\begin{longtable}{llrrcccccll}
\caption{\ha\ fluxes for 76 objects that have been misclassified as PNe in the literature.  The top section contains 71 fluxes determined from SHASSA, and the bottom section contains eight fluxes from VTSS.  Sh\,2-266 and Ap\,2-1 are in both lists.}
\label{tab:mimic_table}\\

\hline \noalign{\smallskip}
PNG & 
Name & 
RAJ2000 & 
DEJ2000 & 
$R_{\rm \NII}$& 
logF$_{\rm red}$ & 
logF(H$\alpha$)&
$r_{\rm aper}$  & 
~$N_{\rm f}$~&
Type&
Note\\
\noalign{\smallskip}
&
& 
&
&
&
&
&
&
&
&
\\
\hline \noalign{\smallskip}

\endfirsthead
\multicolumn{8}{c}{{\tablename} \thetable{} -- Continued} \\[0.5ex]
\hline \noalign{\smallskip}
PNG & 
Name & 
RAJ2000 & 
DEJ2000 & 
$R_{\rm \NII}$& 
logF$_{\rm red}$ & 
logF(H$\alpha$)&
$r_{\rm aper}$  & 
~$N_{\rm f}$~&
Type&
Note\\
\noalign{\smallskip}
&
&
& 
& 
&
&
&
&
&
&
\\
\hline \noalign{\smallskip}

\endhead
\hline \noalign{\smallskip}
\endfoot

\hline \noalign{\smallskip}
\endlastfoot
125.9$-$47.0&PK 125-47 1&00:59:56.7&15:44:14&0.7&$-$10.67&$-$10.77$\pm$0.04&7.4&2&\HII\ &\\
$\ldots$&Reynolds 1&01:11:06.6&10:21:39&$\ldots$&$-$10.9&$-$10.9$\pm$0.2&31.4&2&\HII\ &4\\  
$\ldots$&Sh 2-266&06:18:45.5&15:16:52&0.7&$-$10.23&$-$10.33$\pm$0.04&2.9&2&\HII\ &\\
$\ldots$&Min 2-62&06:43:48.5&$-$01:08:20&0.3&$-$11.14&$-$11.19$\pm$0.04&1.9&4&\HII\ &\\
$\ldots$&MWC 819&06:44:37.7&01:19:32&$\ldots$&$-$11.32&$\dots$&1.8&4&ELS&  1\\
$\ldots$&PHR J0712-0950&07:12:34.8&$-$09:50:44&0.4&$-$9.7&$-$9.7$\pm$0.2&9.5&3&\HII\ &\\  
$\ldots$&EGB 9&07:18:57.9&07:22:23&0.4&$-$11.18&$-$11.25$\pm$0.06&5.4&1&\HII\ &\\
211.9+22.6&EGB 5&08:11:12.8&10:57:17&0.4&$-$11.59&$-$11.65$\pm$0.08&2.8&1&\HII\ &4\\
\end{longtable}
}
\end{center}
{\flushleft  Notes to Table~\ref{tab:mimic_table} follow the format of Table~\ref{tab:shassatable} and Table~\ref{tab:vtss_table}; (C) specific comment given here:  Wray~16-20 --- part of the Vela SNR; H~2-12 --- part of Kepler's SNR; Th~4-4 --- flux is variable;  GK~Per --- flux includes part of outer bipolar nebula.\\ }


\newpage

\noindent{\bf Supplementary Online Tables:}

\begin{center}
{\footnotesize
\begin{longtable}{llrrccrrccl}
\caption{ {\bf (Table 1):}  \ha\ fluxes for 1120 true and possible PNe measured from SHASSA}
\\

\hline \noalign{\smallskip}
PNG & 
Name & 
RAJ2000 & 
DEJ2000 & 
$R_{\rm \NII}$ & 
logF$_{\rm red}$ & 
logF(H$\alpha$)&
$r_{\rm aper}$  & 
$N_{\rm f}$&
c$_{\beta}$&
Note\\
\noalign{\smallskip}
&
& 
& 
& 
&
&
&
&
&
\\
\hline \noalign{\smallskip}

\endfirsthead
\multicolumn{8}{c}{{\tablename} \thetable{} -- Continued} \\[0.5ex]
\hline \noalign{\smallskip}
PNG & 
Name & 
RAJ2000 & 
DEJ2000 & 
$R_{\rm \NII}$& 
logF$_{\rm red}$ & 
logF(H$\alpha$)&
$r_{\rm aper}$  & 
$N_{\rm f}$&
c$_{\beta}$&
Note\\
\noalign{\smallskip}
&
&
& 
& 
& 
&
&
&
&
&
\\
\hline \noalign{\smallskip}

\endhead
\hline \noalign{\smallskip}
\endfoot

\hline \noalign{\smallskip}
\endlastfoot
108.4$-$76.1&BoBn 1&00:37:16.0&$-$13:42:58&0.2&$-$12.03&$-$12.05$\pm$0.06&1.8&3&~0.00~~&C\\
118.8$-$74.7&NGC 246&00:47:03.4&$-$11:52:19&0.0&$-$10.07&$-$10.07$\pm$0.03&6.1&2&0.02&\\
255.3$-$59.6&Lo 1&02:56:58.4&$-$44:10:18&1.1&$-$10.99&$-$11.14$\pm$0.06&4.5&4&$\ldots$&\\
220.3$-$53.9&NGC 1360&03:33:14.6&$-$25:52:18&0.0&$-$9.78&$-$9.78$\pm$0.03&9.4&2&0.00&6\\
206.4$-$40.5&NGC 1535&04:14:15.8&$-$12:44:22&0.0&$-$9.95&$-$9.95$\pm$0.03&5.0&1&0.04&\\
$\ldots$&Fr 2-3&04:56:20.0&$-$28:07:48&1.6&$-$10.89&$-$11.09$\pm$0.12&9.5&1&$\ldots$&1\\ 
243.8$-$37.1&PRTM 1&05:03:01.7&$-$39:45:44&0.0&$-$12.02&$-$12.02$\pm$0.07&1.8&3&$\ldots$&\\  
215.5$-$30.8&Abell 7&05:03:07.5&$-$15:36:23&0.8&$-$10.30&$-$10.41$\pm$0.06&11.0&1&$\ldots$&\\
\end{longtable}
}
\end{center}
\begin{flushleft}
{\scriptsize Notes:  (1) Possible PN; (2) pre-PN; (3) transition object; (4) uncertain counts; (5) confused with nearby object; (6) bad pixels in aperture; (7) object near field edge; (8) flux excludes halo; (9) flux corrected for CSPN;  (10) Wolf-Rayet CSPN; (N) previously unpublished object; (V) very low excitation PN; (C) specific comment given: BoBn~1 --- possibly related to Sgr dSph tidal stream; Te~11 --- possible CV bowshock nebula; Abell~12 is confused with a bright star; K~2-2 ---  flux is for bright inner region only; KLSS~1-8 ---  $R_{\NII}$ is uncertain; HFG~2 ---  flux includes superimposed \HII\ region; Hen~2-25 --- probable symbiotic outflow; Abell~33 ---  nebula is confused with nearby star;  Lo~4 --- variable emission-line central star; Abell~35 ---  unlikely PN (Frew 2008); Mu~1 --- discovered by A. Murrell;  Cn~1-1 ---  yellow symbiotic star;  Mz\,3 --- probable symbiotic outflow;  Abell 38 --- $R_{\NII}$ is uncertain;  M~2-9 --- probable symbiotic outflow;  Terz N 2337 ---  $R_{\NII}$ is very uncertain; PHR~J1757-1649 ---  flux includes superimposed \HII\ region; SB~17 ---  central star is V348~Sgr, a hot R~CrB star; StWr 2-21, Wray~16-423 and Hen 2-436 are Sgr dSph members; Abell~58 ---  born-again object;  NGC~7293 ---  total flux including outer halo is log$F$(\ha) = $-$8.84.
\\}
\end{flushleft}


\newpage

\begin{center}
{\footnotesize
\begin{longtable}{llrrccccl}
\caption{ {\bf (Table 4):}  \ha\ fluxes for 178 true and possible PNe measured from VTSS}
\\
\hline \noalign{\smallskip}
PNG & 
Name & 
RAJ2000 & 
DEJ2000 & 
logF(H$\alpha$)&
$r_{\rm aper}$  & 
$N_{\rm f}$&
c$_{\beta}$&
Note\\
\noalign{\smallskip}
&
& 
&
&
&
&
&
&
\\
\hline \noalign{\smallskip}

\endfirsthead
\multicolumn{8}{c}{{\tablename} \thetable{} -- Continued} \\[0.5ex]
\hline \noalign{\smallskip}
PNG & 
Name & 
RAJ2000 & 
DEJ2000 & 
logF(H$\alpha$)&
$r_{\rm aper}$  & 
$N_{\rm f}$&
c$_{\beta}$&
Note\\
\noalign{\smallskip}
&
&
& 
& 
&
&
& 
&
\\
\hline \noalign{\smallskip}

\endhead
\hline \noalign{\smallskip}
\endfoot

\hline \noalign{\smallskip}
\endlastfoot
118.0$-$08.6&Vy 1-1&00:18:42.2&53:52:20&$-$10.95$\pm$0.09&4.8&1&~0.38~~&6,10?\\
119.3+00.3&BV 5-1&00:20:00.5&62:59:03&$-$11.62$\pm$0.18&5.0&1&$\ldots$&4,5\\  
119.6$-$06.1&Hu 1-1&00:28:15.6&55:57:55&$-$11.00$\pm$0.08&3.2&1&0.44&\\
121.6+00.0&BV 5-2&00:40:21.6&62:51:34&$-$11.60$\pm$0.14&3.2&1&$\ldots$&1,4\\
122.1$-$04.9&Abell 2&00:45:34.7&57:57:35&$-$11.61$\pm$0.10&3.2&2&$\ldots$&6\\
124.3$-$07.7&WeSb 1&01:00:53.3&55:03:48&$-$12.1$\pm$0.2&9.3&1&$\ldots$&\\
126.3+02.9&K 3-90&01:24:58.6&65:38:36&$-$11.90$\pm$0.12&3.2&1&$\ldots$&5,6\\
130.3$-$11.7&M 1-1&01:37:19.4&50:28:12&$-$11.33$\pm$0.10&3.2&1&$\ldots$&\\
130.9$-$10.5&NGC 650/1&01:42:20.0&51:34:31&$-$10.19$\pm$0.05&3.2&1&0.10&\\
138.8+02.8&IC 289&03:10:19.3&61:19:01&$-$10.82$\pm$0.07&3.2&1&1.29&\\
\noalign{\smallskip}
$\ldots$&Fr 2-23&03:14:46.0&48:12:06&$-$10.15$\pm$0.12&31.5&1&$\ldots$&1,N\\  
149.4$-$09.2&HaWe 4&03:27:15.4&45:24:20&$-$11.45$\pm$0.11&3.2&1&$\ldots$&5\\
147.4$-$02.3&M 1-4&03:41:43.4&52:17:00&$-$11.28$\pm$0.09&3.2&1&$\ldots$&\\
149.7$-$03.3&IsWe 1&03:49:05.9&50:00:15&$-$10.88$\pm$0.10&12.5&1&$\ldots$&\\
171.3$-$25.8&Ba 1&03:53:36.6&19:29:39&$-$11.69$\pm$0.12&3.2&2&$\ldots$&6\\
147.8+04.1&M 2-2&04:13:15.0&56:56:58&$-$11.14$\pm$0.09&3.2&1&$\ldots$&\\
151.4+00.5&K 3-64&04:13:27.3&51:51:01&$-$12.3$\pm$0.2&3.2&2&$\ldots$&4,6\\
167.4$-$09.1&K 3-66&04:36:37.2&33:39:30&$-$11.42$\pm$0.10&3.2&1&$\ldots$&\\
174.2$-$14.6&H 3-29&04:37:23.5&25:02:41&$-$11.63$\pm$0.11&3.2&1&$\ldots$&\\
\noalign{\smallskip}
165.5$-$06.5&K 3-67&04:39:47.9&36:45:43&$-$11.14$\pm$0.09&3.2&2&$\ldots$&\\
166.4$-$06.5&CRL 618&04:42:53.7&36:06:53&$-$11.98$\pm$0.14&3.2&2&$\ldots$&2,4\\
205.8$-$26.7&MaC 2-1&05:03:41.9&$-$06:10:03&$-$12.1$\pm$0.2&3.2&2&$\ldots$&4\\
190.3$-$17.7&J 320&05:05:34.3&10:42:23&$-$10.79$\pm$0.07&4.0&1&0.45&5\\
167.0$-$00.9&Abell 8&05:06:38.4&39:08:11&$-$12.00$\pm$0.14&3.2&1&$\ldots$&4\\
173.7$-$05.8&K 2-1&05:07:09.1&30:49:28&$-$11.03$\pm$0.05&3.2&1&1.01&5,6\\
$\ldots$&IPHASX J0511+3028&05:11:51.3&30:28:14&$-$11.78$\pm$0.13&5.8&1&$\ldots$&1,N\\
215.2$-$24.2&IC 418&05:27:28.2&$-$12:41:50&$-$9.02$\pm$0.04&8.0&1&0.32&\\
178.3$-$02.5&K 3-68&05:31:35.9&28:58:42&$-$11.80$\pm$0.13&3.2&1&$\ldots$&4\\
197.2$-$14.2&Abell 10&05:31:45.5&06:56:02&$-$11.40$\pm$0.14&3.2&1&$\ldots$&\\
\noalign{\smallskip}
193.6$-$09.5&H 3-75&05:40:45.0&12:21:23&$-$11.47$\pm$0.11&3.2&1&1.0:&\\
170.7+04.6&K 3-69&05:41:22.2&39:15:08&$-$12.3$\pm$0.2&3.2&1&$\ldots$&4\\
196.6$-$10.9&NGC 2022&05:42:06.2&09:05:11&$-$10.54$\pm$0.05&4.0&2&0.41&\\
184.0$-$02.1&M 1-5&05:46:50.0&24:22:02&$-$11.19$\pm$0.09&3.2&1&$\ldots$&\\
181.5+00.9&Pu 1&05:52:48.4&28:05:59&$-$12.4$\pm$0.2&3.2&1&$\ldots$&4\\
193.0$-$04.5&KLSS 1-5&05:57:08.0&15:25:31&$-$12.1$\pm$0.2&3.2&1&$\ldots$&4\\
197.4$-$06.4&WDHS 1&05:59:24.8&10:41:41&$-$10.40$\pm$0.14&19.2&1&$\ldots$&\\ 
204.0$-$08.5&Abell 13&06:04:47.9&03:56:36&$-$11.45$\pm$0.11&4&1&$\ldots$&\\
201.9$-$04.6&We 1-4&06:14:33.7&07:34:30&$-$12.2$\pm$0.2&3.2&1&$\ldots$&4\\
221.3$-$12.3&IC 2165&06:21:42.8&$-$12:59:14&$-$10.25$\pm$0.08&4.8&1&0.60&\\
\noalign{\smallskip}
218.9$-$10.7&HDW 5&06:23:37.1&$-$10:13:24&$-$11.10$\pm$0.15&5.6&1&$\ldots$&1,6,V\\
204.8$-$03.5&K 3-72&06:23:54.9&05:30:13&$-$11.67$\pm$0.13&3.2&1&$\ldots$&\\
194.2+02.5&J 900&06:25:57.3&17:47:27&$-$10.50$\pm$0.05&4&1&1.13&\\
170.3+15.8&NGC 2242&06:34:07.4&44:46:38&$-$11.58$\pm$0.11&3.2&1&$\ldots$&\\
189.8+07.7&M 1-7&06:37:21.0&24:00:35&$-$11.09$\pm$0.08&3.2&1&1.57&\\
153.7+22.8&Abell 16&06:43:55.5&61:47:25&$-$11.68$\pm$0.11&4.8&1&$\ldots$&\\
224.3$-$05.5&PHR J0652-1240&06:52:20.3&$-$12:40:34&$-$11.62$\pm$0.12&3.2&1&$\ldots$&1\\
204.1+04.7&K 2-2&06:52:23.2&09:57:56&$-$10.07$\pm$0.08&10.4&1&$\ldots$&1,C\\ 
210.3+01.9&M 1-8&06:53:33.8&03:08:27&$-$11.73$\pm$0.13&3.2&1&$\ldots$&\\
221.0$-$01.4&PHR J0701-0749&07:01:09.3&$-$07:49:21&$-$11.79$\pm$0.13&3.2&1&$\ldots$&4\\
\noalign{\smallskip}
226.4$-$03.7&PB 1&07:02:46.8&$-$13:42:35&$-$11.9$\pm$0.3&3.2&1&$\ldots$&4\\
212.0+04.3&M 1-9&07:05:19.2&02:46:59&$-$10.92$\pm$0.09&4.8&1&$\ldots$&\\
217.4+02.0&St 3-1&07:06:50.9&$-$03:05:10&$-$11.61$\pm$0.13&3.2&1&$\ldots$&\\
215.6+03.6&NGC 2346&07:09:22.6&00:48:23&$-$10.55$\pm$0.10&4&1&0.57&9\\
224.9+01.0&We 1-6&07:17:26.0&$-$10:10:38&$-$11.38$\pm$0.15&3.2&1&$\ldots$&\\
227.1+00.5&PHR J0719-1222&07:19:46.7&$-$12:22:47&$-$11.54$\pm$0.13&3.2&1&$\ldots$&5\\
222.1+03.9&PFP 1&07:22:17.7&$-$06:21:46&$-$10.74$\pm$0.14&11.0&1&$\ldots$&\\
214.9+07.8&Abell 20&07:22:57.7&01:45:33&$-$11.65$\pm$0.12&3.2&1&$\ldots$&\\
221.7+05.3&M 3-3&07:26:34.2&$-$05:21:52&$-$11.81$\pm$0.12&3.2&1&$\ldots$&\\
226.7+05.6&M 1-16&07:37:19.0&$-$09:38:50&$-$11.24$\pm$0.09&3.2&1&$\ldots$&\\
\noalign{\smallskip}
228.8+05.3&M 1-17&07:40:22.2&$-$11:32:30&$-$11.28$\pm$0.10&3.2&1&$\ldots$&\\
231.8+04.1&NGC 2438&07:41:50.5&$-$14:44:08&$-$10.32$\pm$0.05&3.2&1&0.80&\\
231.4+04.3&M 1-18&07:42:04.2&$-$14:21:13&$-$11.45$\pm$0.11&3.2&1&$\ldots$&5\\
$\ldots$&Fr 2-25&08:04:04.4&$-$06:30:57&$-$11.10$\pm$0.10&8.4&1&$\ldots$&1,4,N\\
219.1+31.2&Abell 31&08:54:13.2&08:53:53&$-$10.16$\pm$0.08&12.0&1&$\ldots$&7\\ 
016.1+07.7&PTB 20&17:52:15.0&$-$11:10:37&$-$12.2$\pm$0.2&3.2&1&$\ldots$&\\
053.3+24.0&Vy 1-2&17:54:23.0&27:59:58&$-$11.07$\pm$0.08&3.2&1&0.01&\\
014.0+04.8&PTB 19&17:58:25.9&$-$14:25:25&$-$12.4$\pm$0.3&3.2&1&$\ldots$&4\\
096.4+29.9&NGC 6543&17:58:33.4&66:37:59&$-$9.10$\pm$0.04&6.5&1&$\ldots$&\\ 
014.2+03.8&PM 1-205&18:02:38.2&$-$14:42:05&$-$12.1$\pm$0.2&3.2&1&$\ldots$&4\\
\noalign{\smallskip}
019.8+05.6&CTS 1&18:06:59.8&$-$08:55:33&$-$11.84$\pm$0.13&3.2&1&$\ldots$&\\
015.5+02.8&BMP J1808-1406&18:08:35.1&$-$14:06:43&$-$11.70$\pm$0.14&3.2&1&$\ldots$&\\
022.5+04.8&MA 2&18:15:13.4&$-$06:57:12&$-$12.03$\pm$0.15&3.2&1&$\ldots$&4\\
023.0+04.3&MA 3&18:17:49.4&$-$06:48:22&$-$12.05$\pm$0.14&3.2&1&$\ldots$&4\\
021.9+02.7&MaC 1-12&18:21:21.1&$-$08:31:42&$-$11.65$\pm$0.12&3.2&1&$\ldots$&\\
020.6+01.9&PHR J1821-1001&18:21:40.6&$-$10:01:44&$-$11.38$\pm$0.10&3.2&1&$\ldots$&1\\
094.0+27.4&K 1-16&18:21:52.2&64:21:54&$-$11.68$\pm$0.12&3.2&1&$\ldots$&\\
044.3+10.4&We 3-1&18:34:02.3&14:49:10&$-$11.47$\pm$0.10&3.2&1&$\ldots$&\\
042.0+05.4&K 3-14&18:48:32.8&10:35:51&$-$11.42$\pm$0.10&3.2&1&$\ldots$&5\\
051.4+09.6&Hu 2-1&18:49:47.6&20:50:39&$-$10.15$\pm$0.08&4.8&1&0.60&\\
\noalign{\smallskip}
041.8+04.4&K 3-15&18:51:41.5&09:54:53&$-$11.57$\pm$0.12&3.2&1&$\ldots$&V\\
044.0+05.2&K 3-16&18:53:01.6&12:15:59&$-$11.91$\pm$0.14&3.2&1&$\ldots$&4\\
063.1+13.9&NGC 6720&18:53:35.1&33:01:45&$-$9.56$\pm$0.07&4.8&1&0.19&\\
038.7+01.9&YM 16&18:54:57.3&06:02:31&$-$11.30$\pm$0.10&6.4&1&$\ldots$&\\
039.8+02.1&K 3-17&18:56:18.2&07:07:26&$-$11.80$\pm$0.14&3.2&1&$\ldots$&4\\
043.1+03.8&M 1-65&18:56:33.6&10:52:10&$-$11.31$\pm$0.10&3.2&1&$\ldots$&\\
068.7+14.8&Sp 4-1&19:00:26.5&38:21:07&$-$11.37$\pm$0.10&3.2&1&$\ldots$&\\
035.9$-$01.1&Sh 2-71&19:01:59.3&02:09:18&$-$10.82$\pm$0.05&7.2&1&1.01&\\
046.8+03.8&Sh 2-78&19:03:10.1&14:06:59&$-$10.61$\pm$0.09&6.4&1&$\ldots$&\\
051.5+06.1&K 1-17&19:03:37.4&19:21:23&$-$11.82$\pm$0.13&3.2&1&$\ldots$&\\
\noalign{\smallskip}
050.4+05.2&Abell 52&19:04:32.3&17:57:07&$-$11.93$\pm$0.14&3.2&1&$\ldots$&\\
048.5+04.2&K 4-16&19:04:51.5&15:47:38&$-$12.0$\pm$0.2&3.2&1&$\ldots$&4,5,6\\
$\ldots$&IPHASX J1905+1613&19:05:12.4&16:13:47&$-$12.2$\pm$0.2&3.2&1&$\ldots$&1\\
044.1+01.5&PM 1-281&19:06:32.2&10:43:24&$-$11.96$\pm$0.14&3.2&1&$\ldots$&\\
040.3$-$00.4&Abell 53&19:06:45.9&06:23:52&$-$11.66$\pm$0.12&3.2&1&$\ldots$&\\
055.3+06.6&Abell 54&19:08:39.6&22:58:58&$-$12.1$\pm$0.2&3.2&1&$\ldots$&4\\
062.4+09.5&NGC 6765&19:11:06.5&30:32:43&$-$11.30$\pm$0.07&8.0&1&0.40&\\
035.6$-$04.2&MPA J1911+0027&19:11:24.8&00:27:45&$-$11.79$\pm$0.13&3.2&1&$\ldots$&\\
049.4+02.4&Hen 2-428&19:13:05.2&15:46:40&$-$11.39$\pm$0.10&3.2&1&$\ldots$&\\
037.9$-$03.4&Abell 56&19:13:06.1&02:52:48&$-$11.56$\pm$0.14&5.1 &1&$\ldots$&\\
\noalign{\smallskip}
039.5$-$02.7&M 2-47&19:13:34.6&04:38:04&$-$11.36$\pm$0.11&3.2&1&$\ldots$&\\
048.7+01.9&Hen 2-429&19:13:38.4&14:59:19&$-$11.25$\pm$0.09&3.2&1&$\ldots$&10\\
038.7$-$03.3&M 1-69&19:13:54.0&03:37:42&$-$11.24$\pm$0.10&4.0&1&$\ldots$&\\
051.0+03.0&Hen 2-430&19:14:04.2&17:31:33&$-$11.41$\pm$0.10&3.2&1&$\ldots$&\\
051.0+02.8&WhMe 1&19:14:59.8&17:22:46&$-$12.11$\pm$0.13&3.2&1&$\ldots$&1,4,5\\
040.4$-$03.1&K 3-30&19:16:27.7&05:13:19&$-$11.56$\pm$0.12&3.2&1&$\ldots$&\\
058.6+06.1&Abell 57&19:17:05.7&25:37:33&$-$11.86$\pm$0.13&3.2&1&$\ldots$&\\
041.8$-$02.9&NGC 6781&19:18:28.1&06:32:19&$-$10.01$\pm$0.08&5.6&1&1.11&\\
052.9+02.7&K 3-31&19:19:02.7&19:02:21&$-$12.04$\pm$0.14&4.0&1&$\ldots$&6\\
077.6+14.7&Abell 61&19:19:10.2&46:14:52&$-$11.38$\pm$0.05&7.7&1&$\ldots$&\\
\noalign{\smallskip}
051.3+01.8&PM 1-295&19:19:18.8&17:11:48&$-$12.01$\pm$0.14&3.2&1&$\ldots$&4\\
076.3+14.1&Pa 5        & 19:19:30.5&44:45:43&$-$11.56$\pm$0.12&3.7&1&$\ldots$&\\
043.0$-$03.0&M 4-14&19:21:00.7&07:36:52&$-$11.73$\pm$0.13&3.2&1&$\ldots$&\\
055.3+02.7&He 1-1&19:23:46.9&21:06:39&$-$12.1$\pm$0.2&3.2&3&$\ldots$&6\\
056.0+02.0&K 3-35&19:27:44.0&21:30:04&$-$12.4$\pm$0.2&3.2&1&$\ldots$&3,4\\
061.3+03.6&M 1-91&19:32:57.7&26:52:43&$-$11.74$\pm$0.12&3.2&1&$\ldots$&1\\
059.4+02.3&K 3-37&19:33:46.8&24:32:27&$-$12.1$\pm$0.2&3.2&1&$\ldots$&4\\
064.7+05.0&BD +30 3639&19:34:45.2&30:30:59&$-$9.41$\pm$0.06&4.8&1&0.51&10,V\\
059.9+02.0&K 3-39&19:35:54.5&24:54:48&$-$12.3$\pm$0.2&3.2&1&$\ldots$&4\\
055.5$-$00.5&M 1-71&19:36:26.9&19:42:24&$-$10.96$\pm$0.08&4.0&1&2.44&\\
\noalign{\smallskip}
060.5+01.8&Hen 2-440&19:38:08.4&25:15:41&$-$11.73$\pm$0.12&3.2&1&$\ldots$&\\
052.5$-$02.9&Me 1-1&19:39:09.8&15:56:48&$-$10.81$\pm$0.07&4.0&1&0.38&\\
061.8+02.1&Hen 2-442&19:39:43.4&26:29:33&$-$11.57$\pm$0.12&3.2&1&$\ldots$&\\
051.9$-$03.8&M 1-73&19:41:09.3&14:56:59&$-$10.84$\pm$0.07&3.2&1&$\ldots$&\\
054.4$-$02.5&M 1-72&19:41:34.0&17:45:18&$-$11.20$\pm$0.09&3.2&1&$\ldots$&\\
053.8$-$03.0&Abell 63&19:42:10.4&17:05:15&$-$11.47$\pm$0.11&3.2&1&$\ldots$&\\
052.2$-$04.0&M 1-74&19:42:18.9&15:09:08&$-$10.98$\pm$0.08&4.0&1&0.98&6\\
054.2$-$03.4&IPHASX J1943-1709&19:43:59.5&17:09:01&$-$11.88$\pm$0.14&3.2&1&$\ldots$&4\\
059.1$-$00.7&Kn 9 &19:44:59.0&22:45:48&$-$11.10$\pm$0.11&4.5&2&$\ldots$&1\\
057.9$-$01.5&Hen 2-447&19:45:22.2&21:20:04&$-$11.61$\pm$0.12&3.2&1&$\ldots$&\\
\noalign{\smallskip}
043.5$-$13.4&Abell 67&19:58:27.0&03:02:52&$-$11.90$\pm$0.14&3.2&1&$\ldots$&4\\
060.0$-$04.3&Abell 68&20:00:10.6&21:42:55&$-$11.61$\pm$0.12&3.2&1&$\ldots$&\\
058.6$-$05.5&WeSb 5&20:01:42.0&19:54:41&$-$11.32$\pm$0.10&3.2&2&$\ldots$&5\\
107.0+21.3&K 1-6&20:04:13.4&74:26:28&$-$11.37$\pm$0.10&4.0&2&$\ldots$&\\
082.1+07.0 &NGC 6884&20:10:23.7&46:27:40&$-$10.57$\pm$0.06&3.2&1&0.26&\\
057.2$-$08.9&NGC 6879&20:10:26.7&16:55:21&$-$11.03$\pm$0.08&3.2&1&0.29&\\
060.3$-$07.3&Hen 1-5&20:11:56.1&20:20:04&$-$11.42$\pm$0.10&3.2&1&$\ldots$&\\
060.1$-$07.7&NGC 6886&20:12:42.8&19:59:23&$-$10.63$\pm$0.09&4.8&1&$\ldots$&\\
066.9$-$05.2&PC 24&20:19:38.1&27:00:11&$-$11.37$\pm$0.10&3.2&1&$\ldots$&\\
058.3$-$10.9&IC 4997&20:20:08.7&16:43:54&$-$9.92$\pm$0.08&7.2&1&$\ldots$&6\\
\noalign{\smallskip}
061.4$-$09.5&NGC 6905&20:22:22.9&20:06:17&$-$10.48$\pm$0.05&4.0&1&0.00&10\\
$\ldots$&PM 1-329&20:50:13.6&59:45:51&$-$12.2$\pm$0.2&3.2&1&$\ldots$&1,4\\
101.6+13.0&Kn 49&20:55:48.0&65:34:00&$-$11.67$\pm$0.12&3.2&2&$\ldots$&1,C\\
089.0+00.3&NGC 7026&21:06:18.2&47:51:05&$-$10.16$\pm$0.08&4.8&1&0.88&10\\
084.9$-$03.4&NGC 7027&21:07:01.7&42:14:10&$-$9.29$\pm$0.06&6.4&2&1.20&\\
084.2$-$04.2&K 3-80&21:07:39.7&40:57:52&$-$11.80$\pm$0.13&3.2&1&$\ldots$&1,4\\  
082.1$-$07.8&Kn 24&21:13:37.7&37:15:37&$-$11.29$\pm$0.11&4.0&1&$\ldots$&\\
089.8$-$00.6&Sh 1-89&21:14:07.6&47:46:22&$-$11.38$\pm$0.11&3.2&1&$\ldots$&\\
088.7$-$01.6&NGC 7048&21:14:15.2&46:17:18&$-$10.85$\pm$0.06&7.6&1&$\ldots$&\\
072.7$-$17.1&Abell 74&21:16:52.3&24:08:52&$-$10.76$\pm$0.09&7.2&2&$\ldots$&\\
\noalign{\smallskip}
080.3$-$10.4&MWP 1&21:17:08.3&34:12:27&$-$10.74$\pm$0.10&4.8&1&$\ldots$&7\\ 
089.3$-$02.2&M 1-77&21:19:07.4&46:18:47&$-$11.21$\pm$0.09&3.2&1&$\ldots$&V\\
093.9$-$00.1&IRAS 21282+5050&21:29:58.1&51:04:00&$-$11.71$\pm$0.16&3.2&1&$\ldots$&3,4,10\\
096.3+02.3&K 3-61&21:30:00.7&54:27:27&$-$11.87$\pm$0.14&3.2&1&$\ldots$&4,10\\
093.3$-$00.9&K 3-82&21:30:51.6&50:00:07&$-$11.7$\pm$0.2&3.2&2&$\ldots$&\\
089.8$-$05.1&IC 5117&21:32:31.0&44:35:49&$-$10.55$\pm$0.05&4.0&1&1.16&\\
086.5$-$08.8&Hu 1-2&21:33:08.4&39:38:10&$-$10.57$\pm$0.09&4.8&1&0.57&\\
094.5$-$00.8A&LeDu 1&21:36:05.5&50:54:10&$-$12.2$\pm$0.3&5.0&1&$\ldots$&4\\
066.7$-$28.2&NGC 7094&21:36:53.0&12:47:19&$-$11.24$\pm$0.09&3.2&1&$\ldots$&\\
093.3$-$02.4&M 1-79&21:37:01.5&48:56:03&$-$10.95$\pm$0.08&3.2&2&1.01&\\
\noalign{\smallskip}
095.0$-$05.5&GLMP 1047&21:56:32.9&47:36:13&$-$11.82$\pm$0.14&3.2&1&$\ldots$&3\\
103.2+00.6&M 2-51&22:16:03.9&57:28:34&$-$11.14$\pm$0.09&3.2&1&$\ldots$&\\
104.1+01.0&Bl 2-1&22:20:16.6&58:14:17&$-$11.55$\pm$0.14&3.2&1&$\ldots$&4\\
103.7+00.4&M 2-52&22:20:30.8&57:36:22&$-$11.69$\pm$0.14&3.2&1&$\ldots$&4\\
100.6$-$05.4&IC 5217&22:23:55.7&50:58:00&$-$10.64$\pm$0.06&4.0&2&0.01&\\
102.9$-$02.3&Abell 79&22:26:17.3&54:49:38&$-$11.20$\pm$0.12&3.2&1&$\ldots$&9\\
099.7$-$08.8&HaWe 15&22:30:33.4&47:31:24&$-$11.29$\pm$0.10&4.0&1&$\ldots$&1\\
100.0$-$08.7&Me 2-2&22:31:43.7&47:48:04&$-$10.68$\pm$0.06&4.0&1&$\ldots$&\\
104.4$-$01.6&M 2-53&22:32:17.7&56:10:26&$-$11.45$\pm$0.11&3.2&1&$\ldots$&6\\
102.8$-$05.0&Abell 80&22:34:45.6&52:26:06&$-$11.45$\pm$0.11&3.2&2&$\ldots$&\\
\noalign{\smallskip}
107.8+02.3&NGC 7354&22:40:19.9&61:17:08&$-$10.49$\pm$0.05&3.2&1&1.98&\\
104.8$-$06.7&M 2-54&22:51:38.9&51:50:43&$-$11.20$\pm$0.09&3.2&2&$\ldots$&1,3\\
107.7$-$02.2&M 1-80&22:56:19.8&57:09:21&$-$11.50$\pm$0.13&3.2&1&$\ldots$&\\
104.2$-$29.6&Jones 1&23:35:53.3&30:28:06&$-$10.82$\pm$0.09&4.8&1&$\ldots$&7\\
110.6$-$12.9&K 1-20&23:39:10.8&48:12:29&$-$12.2$\pm$0.2&3.2&1&$\ldots$&\\
114.0$-$04.6&Abell 82&23:45:47.8&57:03:59&$-$11.37$\pm$0.10&3.2&2&$\ldots$&\\
112.9$-$10.2&Abell 84&23:47:44.3&51:23:56&$-$11.13$\pm$0.09&3.2&1&$\ldots$&\\
116.2+08.5&M 2-55&23:31:52.7&70:22:10&$-$11.21$\pm$0.09&3.2&1&$\ldots$&\\

\end{longtable}
}
\end{center}
\begin{flushleft}
{\scriptsize Notes:  (1) Possible PN; (2) pre-PN; (3) transition object; (4) uncertain counts; (5) confused with nearby object; (6) bad pixels in aperture; (7) object near field edge; (8) flux excludes halo; (9) flux corrected for CSPN;  (10) Wolf-Rayet CSPN; (N) previously unpublished object; (V) very low excitation PN; (C) specific comment given:  K~2-2 --- flux is for bright inner region only; Kn~49 --- possibly an isolated SNR filament.
\\}
\end{flushleft}


\label{lastpage}
\end{document}